\input harvmac
\input epsf.tex
\input amssym.tex
\overfullrule=0pt

%
\def\tilde{\widetilde}

\def\S{{\bf S}}

\font\zfont = cmss10 

\def\bigone{\hbox{1\kern -.23em {\rm l}}}
\def\ZZ{\hbox{\zfont Z\kern-.4emZ}}

\noblackbox
 

\def\lem#1{\bigskip\noindent{\bf Lemma #1} } 
\def\prop#1{\bigskip\noindent{\bf Proposition #1} } 
 
\def\rmk#1{\bigskip\noindent{\bf Remark #1.} } 
%


\def\figin{\epsfcheck\figin}\def\figins{\epsfcheck\figins}
\def\epsfcheck{\ifx\epsfbox\UnDeFiNeD
\message{(NO epsf.tex, FIGURES WILL BE IGNORED)}
\gdef\figin##1{\vskip2in}\gdef\figins##1{\hskip.5in}
\else\message{(FIGURES WILL BE INCLUDED)}%
\gdef\figin##1{##1}\gdef\figins##1{##1}\fi}
\def\DefWarn#1{}
\def\figinsert{\goodbreak\midinsert}
\def\ifig#1#2#3{\DefWarn#1\xdef#1{fig.~\the\figno}
\writedef{#1\leftbracket fig.\noexpand~\the\figno}%
\figinsert\figin{\centerline{#3}}\medskip\centerline{\vbox{\baselineskip12pt
\advance\hsize by -1truein\noindent\footnotefont{\bf Fig.~\the\figno:} #2}}
\bigskip\endinsert\global\advance\figno by1}


\def\unlockat{\catcode`\@=11}
\def\lockat{\catcode`\@=12}

\unlockat

\def\newsec#1{\global\advance\secno by1\message{(\the\secno. #1)}
\global\subsecno=0\global\subsubsecno=0\eqnres@t\noindent
{\bf\the\secno. #1}
\writetoca{{\secsym} {#1}}\par\nobreak\medskip\nobreak}
\global\newcount\subsecno \global\subsecno=0
\def\subsec#1{\global\advance\subsecno
by1\message{(\secsym\the\subsecno. #1)}
\ifnum\lastpenalty>9000\else\bigbreak\fi\global\subsubsecno=0
\noindent{\it\secsym\the\subsecno. #1}
\writetoca{\string\quad {\secsym\the\subsecno.} {#1}}
\par\nobreak\medskip\nobreak}
\global\newcount\subsubsecno \global\subsubsecno=0
\def\subsubsec#1{\global\advance\subsubsecno by1
\message{(\secsym\the\subsecno.\the\subsubsecno. #1)}
\ifnum\lastpenalty>9000\else\bigbreak\fi
\noindent\quad{\secsym\the\subsecno.\the\subsubsecno.}{#1}
\writetoca{\string\qquad{\secsym\the\subsecno.\the\subsubsecno.}{#1}}
\par\nobreak\medskip\nobreak}

\def\subsubseclab#1{\DefWarn#1\xdef
#1{\noexpand\hyperref{}{subsubsection}%
{\secsym\the\subsecno.\the\subsubsecno}%
{\secsym\the\subsecno.\the\subsubsecno}}%
\writedef{#1\leftbracket#1}\wrlabeL{#1=#1}}
\lockat



%

\def\IL{\relax{\rm I\kern-.18em L}}
\def\IH{\relax{\rm I\kern-.18em H}}
\def\IR{\relax{\rm I\kern-.18em R}}
\def\IC{\relax\hbox{$\inbar\kern-.3em{\rm C}$}}
\def\IZ{\relax\ifmmode\mathchoice
{\hbox{\cmss Z\kern-.4em Z}}{\hbox{\cmss Z\kern-.4em Z}}
{\lower.9pt\hbox{\cmsss Z\kern-.4em Z}}
{\lower1.2pt\hbox{\cmsss Z\kern-.4em Z}}\else{\cmss Z\kern-.4em Z}\fi}

\font\manual=manfnt \def\dbend{\lower3.5pt\hbox{\manual\char127}}

\def\IZ{\relax\ifmmode\mathchoice
{\hbox{\cmss Z\kern-.4em Z}}{\hbox{\cmss Z\kern-.4em Z}}
{\lower.9pt\hbox{\cmsss Z\kern-.4em Z}}
{\lower1.2pt\hbox{\cmsss Z\kern-.4em Z}}\else{\cmss Z\kern-.4em Z}\fi}


\def\IZ{\relax\ifmmode\mathchoice
{\hbox{\cmss Z\kern-.4em Z}}{\hbox{\cmss Z\kern-.4em Z}}
{\lower.9pt\hbox{\cmsss Z\kern-.4em Z}}
{\lower1.2pt\hbox{\cmsss Z\kern-.4em Z}}\else{\cmss Z\kern-.4em
Z}\fi}
\def\IB{\relax{\rm I\kern-.18em B}}
\def\IC{{\relax\hbox{$\inbar\kern-.3em{\rm C}$}}}
\def\ID{\relax{\rm I\kern-.18em D}}
\def\IE{\relax{\rm I\kern-.18em E}}
\def\IF{\relax{\rm I\kern-.18em F}}
\def\IG{\relax\hbox{$\inbar\kern-.3em{\rm G}$}}
\def\IGa{\relax\hbox{${\rm I}\kern-.18em\Gamma$}}
\def\IH{\relax{\rm I\kern-.18em H}}
\def\II{\relax{\rm I\kern-.18em I}}
\def\IK{\relax{\rm I\kern-.18em K}}
\def\IP{\relax{\rm I\kern-.18em P}}

\def\IQ{\relax\hbox{$\inbar\kern-.3em{\rm Q}$}}
\def\IP{\relax{\rm I\kern-.18em P}}

\def\IB{\relax{\rm I\kern-.18em B}}
\def\IC{\Bbb{C} }
\def\ID{\relax{\rm I\kern-.18em D}}
\def\IE{\relax{\rm I\kern-.18em E}}
\def\IF{\relax{\rm I\kern-.18em F}}
\def\IG{\relax\hbox{$\inbar\kern-.3em{\rm G}$}}
\def\IGa{\relax\hbox{${\rm I}\kern-.18em\Gamma$}}
\def\IH{\relax{\rm I\kern-.18em H}}
\def\II{\relax{\rm I\kern-.18em I}}
\def\IJ{\relax{\rm I\kern-.18em J}}
\def\IK{\relax{\rm I\kern-.18em K}}
\def\IL{\relax{\rm I\kern-.18em L}}

\def\IN{\relax{\rm I\kern-.18em N}}
\def\IO{\relax{\rm I\kern-.18em O}}
\def\IP{\relax{\rm I\kern-.18em P}}
\def\IQ{\relax\hbox{$\inbar\kern-.3em{\rm Q}$}}
\def\IR{\relax{\rm I\kern-.18em R}}
\def\IW{\relax\hbox{$\inbar\kern-.3em{\rm W}$}}

\def\inbar{\,\vrule height1.5ex width.4pt depth0pt}

\font\cmss=cmss10 \font\cmsss=cmss10 at 7pt
\def\IR{\relax{\rm I\kern-.18em R}}

\def\IC{{\relax\hbox{$\inbar\kern-.3em{\rm C}$}}}

\Title{ } {\vbox{\centerline{Constraining the K\" ahler Moduli}
\smallskip
\centerline{in the}
\smallskip
\centerline{Heterotic Standard Model}
}}
\smallskip
\centerline{Tom\'as L. G\'omez$^{\, a}$, Sergio Lukic$^{\, b}$ and Ignacio Sols$^{\, c}$}
\smallskip
\centerline{$^a${\it CSIC Instituto de Matem\'aticas y Fisica Fundamental}}
\centerline{\it Madrid 28006, Spain.}
\centerline{\tt tg@imaff.cfmac.csic.es}
\centerline{$^b${\it Department of Physics and Astronomy, Rutgers University}}
\centerline{\it Piscataway, NJ 08855-0849, USA.}
\centerline{\tt lukic@physics.rutgers.edu}
\centerline{$^c${\it Departamento de \'Algebra, Facultad de Matem\'aticas UCM}}
\centerline{\it Madrid 28040, Spain.}
\centerline{\tt isols@mat.ucm.es}

\bigskip

Phenomenological implications of the volume of the Calabi-Yau threefolds
on the hidden and observable M-theory boundaries, together with
slope stability of their corresponding vector bundles,
constrain the set of K\"ahler moduli which give rise to realistic compactifications of the 
strongly coupled heterotic string.
When vector bundles are constructed using extensions, we provide simple rules to 
determine lower and upper bounds to the region of the K\"ahler moduli space 
where such compactifications can exist. 
We show how small these regions can be, working out in full detail the case of the
recently proposed Heterotic Standard Model. More explicitely, we exhibit 
K\"ahler classes in these regions for
which the visible vector bundle is stable. On the other hand, there is no 
polarization for which the hidden bundle is stable.

\bigskip

\noindent

\Date{December, 2005}

\newsec{Introduction}

\nref\banksdine{T. Banks and M. Dine, 
{\it Couplings and Scales in Strongly Coupled Heterotic String Theory},  
{\tt hep-th/9605136} (1996). 
}%
\nref\upennbundle{V. Braun, Y-H. He, B. A. Ovrut and T. Pantev, 
{\it Vector Bundle Extensions, Sheaf Cohomology, and the Heterotic Standard Model},  
{\tt hep-th/0505041} (2005). 
}%
\nref\HSMmoduli{V. Braun, Y-H. He, B. A. Ovrut and T. Pantev, 
{\it Heterotic Standard Model Moduli},  
{\tt hep-th/0509051} (2005). 
}%
\nref\HSMlatest{V. Braun, Y-H. He, B. A. Ovrut and T. Pantev, 
{\it The Exact MSSM Spectrum from String Theory},  
{\tt hep-th/0512177} (2005). 
}%
\nref\ellipticCY{V. Braun, B. A. Ovrut, T. Pantev and R. Reinbacher, 
{\it Elliptic Calabi-Yau Threefolds with $\IZ_{3}\times \IZ_{3}$ Wilson Lines},  
{\tt hep-th/0410055} (2004). 
}%
\nref\curiokrausea{G. Curio and A. Krause, 
Nucl.Phys. {\bf B602} 172-200,  
{\tt hep-th/0012152} (2001). 
}%
\nref\curiokrauseb{G. Curio and A. Krause, 
Nucl.Phys. {\bf B693} 195-222,  
{\tt hep-th/0308202} (2004). 
}%
\nref\DonagiBouchard{R. Donagi and V. Bouchard, 
{\it An SU(5) Heterotic Standard Model},  
{\tt hep-th/0512149} (2005). 
}%
\nref\donaldson{S. K. Donaldson and P. B. Kronheimer,
{\it The Geometry of Four-Manifolds},
Oxford University Press, Oxford, (1990). 
}
\nref\douglas{M. R. Douglas, 
{\it The Statistics of String/M-Theory Vacua },  
JHEP 0305:046,
{\tt hep-th/0303194} (2003). 
}%
\nref\pistable{M. R. Douglas, B. Fiol and C. R\"omelsberger, 
{\it Stability and BPS branes},  
{\tt hep-th/0002037} (2000). 
}%
\nref\automorphismcone{A. Grassi and D. R. Morrison, 
{\it Automorphisms and the K\"ahler cone of certain Calabi-Yau manifolds},  
Duke Math. J. {\bf 71}, 831-838 (1993), {\tt alg-geom/9212004}. 
}%
\nref\gukov{S. Gukov, S. Kachru, X. Liu and L. McAllister, 
{\it Heterotic Moduli Stabilization with Fractional Chern-Simons Invariants},  
{\tt hep-th/0310159} (2003). 
}%
\nref\hartshorne{R. Hartshorne,
{\it Algebraic Geometry},
Springer-Verlag, New York, (1977). Graduate Texts in Mathematics, No. 52. 
}
\nref\HW{P. Ho\v rava and E. Witten, 
{\it Eleven-dimensional supergravity on a manifold with boundary},
{\tt hep-th/9603142} (1996). 
}%
\nref\joyce{D. D. Joyce, 
{\it Compact Manifolds with Special Holonomy},  
Oxford University Press (2000). 
}%
\nref\effectiveclasses{E. Looijenga, 
{\it Rational surfaces with an anti-canonical cycle},  
Ann. of Math. (2) {\bf 114} (1981). 
}%
\nref\namikawa{Yo. Namikawa, 
{\it On the birational structure of certain Calabi-Yau threefolds},  
J. Math. Kyoto Univ. {\bf 31}, 151-164 (1991). 
}%
\nref\S{C. Schoen,
{\it On the fiber products of rational elliptic surfaces with sections},
Math. Ann. {\bf 197}, 177-199 (1988).
}
\nref\sharpe{E. Sharpe, 
{\it K\"ahler Cone Substructure},  
Adv. Theor. Math. Phys. {\bf 2} (1998) 1441, 
{\tt hep-th/9810064}. 
}%
\nref\wilsoncone{P. M. H. Wilson, 
{\it The K\"ahler cone on Calabi-Yau threefolds},  
Invent. Math. {\bf 107}, 561-583 (1992). 
}%
\nref\witten{E. Witten, 
{\it Strong coupling expansion of Calabi-Yau compactification},  
{\tt hep-th/9602070} (1996). 
}%

Our understanding of Calabi-Yau compactifications of string/M-theory has
been increased considerably during the last years.
On the one hand, distributions of vacua for type $\II {\rm B}$,
$\II {\rm A}$ and type ${\rm I}$ string theory are much better understood. On the
other hand, promising compactifications of the heterotic string
have been found at special points of the moduli space.

Although a systematic study of 
distributions of vacua for compactifications of the
heterotic string is
much harder, 
because our primitive understanding of their 
moduli stabilization and the huge amount of vector bundle moduli, 
we can
still find systematic criteria to
constrain the regions of the moduli space
where realistic vacua should be located.

Recently, phenomenologically interesting
Calabi-Yau compactifications of the heterotic string
have appeared in the literature \upennbundle, \ellipticCY.
Using certain elliptically fibered threefold with fundamental group 
$\IZ_{3}\times \IZ_{3}$, and an
$SU(4)\times \IZ_{3}\times \IZ_{3}$ instanton living on
the visible $E_{8}$-bundle, 
give rise to an effective field theory on $\IR^4$ which has the particle spectrum
of the Minimal Supersymmetric Standard Model (MSSM), 
with no exotic matter but an additional pair of Higgs-Higgs conjugate superfields.
In these models, 
vector bundles are constructed using vector
bundle extensions, 
which correspond to Hermitian Yang-Mills connections when they are slope-stable.
We use this specific construction to exemplify how a systematic selection of realistic K\"ahler moduli 
can be done. \foot{Recently, 
Donagi and Bouchard \DonagiBouchard\ have also proposed
an independent CY compactification of the heterotic string 
with the spectrum of the MSSM and no exotic matter, 
using a different Calabi-Yau with an
explicitly slope-stable vector bundle in the observable sector. 
It would be also interesting to study in detail these questions with the  
vector bundle which has just appeared in \HSMlatest, on the same  
CY \ellipticCY.}
  
The organization of the paper is as follows: section 2 contains an outline of the natural
criteria for selecting K\"ahler moduli in realistic Calabi-Yau compactifications 
of the heterotic string.
In section 3, we analyze the case of the Heterotic Standard Model, describe the 
geometry of the elliptic Calabi-Yau and construct its K\"ahler cone. Section 4 provides
lower and upper bounds to the region of the K\"ahler cone that makes stable the
observable vector bundle of the HSM. In such construction we find a destabilizing 
sub-line bundle for the hidden vector bundle, and exhibit K\"ahler classes that make stable
the visible one. 

\newsec{ Picking K\"ahler moduli}

The spacetime in a Calabi-Yau compactification of the strongly coupled heterotic 
string \HW, is defined through the direct product
eleven-dimensional manifold $Y=\IR^4\times X\times [0,\, 1]$,
with $X$ a Calabi-Yau threefold. ${\cal N}=1$ supersymmetry
on the four dimensional Effective Field Theory, requires to 
fix a $G_2$-holonomy metric on $X\times [0,\, 1]$ plus gauge
connections at the hidden and visible vector bundles, 
which satisfy the Hermitian Yang-Mills equations. In order
to define a barely $G_2$-holonomy metric on $X\times [0,\, 1]$ we
introduce a calibration $3$-form, according to D. Joyce \joyce
\eqn\calibration{
\Phi = (At + B)\omega\wedge {\rm d}t + {\rm Re}(\Omega),
}
which depends on the differential ${\rm dt}$ along the interval and 
the holomorphic $3$-form $\Omega$  
and K\"ahler class $\omega$ of the threefold. 
Such a calibration defines a barely $G_2$-holonomy metric 
on $X\times [0,\, 1]$, 
where the K\"ahler class is linearly dilated along the interval,
therefore at the visible and hidden boundaries the K\"ahler classes are 
$\omega_0=B\omega$ and $\omega_1=(A+B)\omega$
($i=1$ stands for
the `hidden' boundary and $i=0$ for the `visible' one).
The set of K\"ahler classes on
$X$ is usually known as the K\"ahler cone and denoted by
${\cal K}(X)\subset H^2 (X,\, \IZ)$.

One approach to model building is to attach a
$SU(n)\times G$ Hermitian Yang-Mills gauge connection at 
the boundary, to obtain 
an Effective Field Theory with the commutant of 
$SU(n)\times G\subset E_8$ as gauge group while the
${\cal N}=1$ supersymmetry of the EFT is preserved. 
Here, $G$ is the 
non-trivial holonomy group associated to certain flat line bundle.
By the theorem of Donaldson and Uhlenbeck-Yau \donaldson,
we know that $SU(n)$-connections that satisfy the 
Hermitian Yang-Mills equations and slope-stable
rank-$n$ holomorphic vector bundles with vanishing
first Chern class, are in one-to-one correspondence.

\subsec{Constraining angular degrees of freedom}

Thus, the holomorphic vector bundles $V_i\to X$, that we fix
at the hidden and visible sectors, have to be slope stable in order to get a sensible vacuum. 
Slope stability can impose severe constraints on the 
K\"ahler moduli. 

If $W_{i}\hookrightarrow V_i$ is a rank-$m$ (with $m<n$)
holomorphic torsion free subsheaf\foot{
It is enough to consider reflexive sheaves, i.e.,
sheaves with $W_i^{}=W_i^{\vee\vee}$. Furthermore, we can assume that
$W_i$ is semistable.}, then only the 
$\omega_{i}\in {\cal K}(X)$ that verify
\eqn\ineqone{
{1\over m} \int_{X}\omega_{i}^2 \wedge c_1 (W_i)<
{1\over n} \int_{X}\omega_{i}^2 \wedge c_1 (V_i) = 0,
}
can make $V_{i}$ stable.
At this point we realize that if $\omega_i$ is stablemaker for $V_i$, 
then $N\omega_i$ with $N\in \IZ^+$ is also stablemaker.
The stablemakers form a subcone 
${\cal K}^{s}_i(X)\subseteq {\cal K}(X)$ within the K\"ahler moduli,
\sharpe. 

The physical importance of slope stability is clear, \donaldson:
Non-stablemaker classes at the boundary of ${\cal K}^{s}_i(X)$ make
the vector bundle $V_i$ semistable, i.e.
we can only find correspondences to
connections with reduced gauge group $H\subset SU(n)$, thus
the gauge dynamics of the Effective Field Theory would be governed 
by the commutant of $H\times G\subset E_8$ instead of $SU(n)\times G$.

Usually, a detailed computation of ${\cal K}^{s}_i(X)$ is difficult
because we need to identify every subsheaf $W_i$ of $V_i$.
Note that, if $h^0(W_i^\vee\otimes V_i)=0$, then $W_i$
cannot be a subsheaf of $V_i$, but the converse is not necessarily 
true.
If the vector bundle $V_i$ is 
constructed through a non-trivial extension,
defined by a short exact sequence
\eqn\defext{
0\longrightarrow V_{L}\longrightarrow V_{i}\longrightarrow V_{R}\longrightarrow 0,
}
with ${\rm Ext}^1 (V_R,\, V_L)\neq 0$, we can give upper and lower bounds
to ${\cal K}^{s}_i(X)$ in a simple way, looking at subsheaves of
$V_L$ and $V_R$.

On the one hand, the set ${\cal UL}_i$ 
of subsheaves of $V_{L}$ is a subset of the
set of subsheaves of $V_{i}$, since $V_L\rightarrow V_{i}$ is injective.
This provides an upper bound for cone ${\cal K}^s_i(X)$ of K\"ahler 
classes for which $V_i$ is stable:
\eqn\upperone{
{\cal K}_i^s(X)^>=\Big\{
\omega_i\in {\cal K}(X)\, :\, \int_X \omega_i^2\wedge c_1(L_i) <0,\,\, \forall\, L_i
\in {\cal UL}_i
\Big\},
}
On the other hand, a subsheaf of $V_i$ gives an element of
${\cal UL}_i \times {\cal UR}_i$, where ${\cal UR}_i$ is the subset
of subsheaves of $V_R$. Indeed, if $W_i$ is a subsheaf of $V_i$,
there is a commutative diagram
\eqn\comdiag{\matrix{
0 & \longrightarrow & V_{L} & \longrightarrow & V_{i} & 
\longrightarrow & V_{R} & \longrightarrow & 0 \cr
 & & \uparrow & & \uparrow & & \uparrow & & \cr
0 & \longrightarrow & W_{L} & \longrightarrow & W_{i} & 
\longrightarrow & W_{R} & \longrightarrow & 0 \cr
}}
where the vertical arrows are injective, hence we obtain subsheaves
$W_L$ and $W_R$ of $V_L$ and $V_R$. This gives a lower bound
\eqn\lowerone{
{\cal K}_i^s(X)^< =\Big\{
\omega_i\in {\cal K}(X)\, :\, \int_X \omega_i^2\wedge 
\big( c_1(W_L)+c_1(W_R)\big)  <0,
\,\, \forall\, W_L\in  {\cal UL}_i\, , \; W_R \in{\cal UR}_i
\Big\}.
}
Note that the ones belonging to ${\cal UL}_i$ are true subsheaves 
of $V_i$, and the ones in
${\cal UR}_i$ are possible subsheaves of $V_i$. Therefore, we can 
construct two bounds to the stablemaker K\"ahler subcone ${\cal K}_i^s(X)$,
\eqn\bounds{
{\cal K}_i^s(X)^< \subseteq {\cal K}_i^s(X)
\subseteq {\cal K}_i^s(X)^>
}
Sometimes, we can use further information to discard some pairs
$(W_L,W_R)$ which do not come from subsheaves $W_i$ of $V_i$, hence
obtaining a better lower bound.
For instance, the pair $(0,V_R)$ can be discarded, because it
would give a splitting of the defining exact sequence \defext, but 
we have assumed that the extension is not trivial, hence has no
splitting. Another cases that can be discarded are pairs of the
form $(0,W_R)$ when $h^0(W_R^\vee\otimes V_i)=0$.
We shall apply these ideas in the next sections,
to the vector bundles constructed in the Heterotic Standard
Model, \upennbundle.

\subsec{Constraining radial degrees of freedom}

In the last subsection 
we have seen how to choose rays in the K\"ahler cone 
that preserve the slope stability of a given vector bundle, and thus define a 
consistent gauge group
in the effective field theory. On the other hand, 
radial degrees of freedom in ${\cal K}^s_i(X)$ are
related with variations of the volume of $X$, \pistable.
We are not free to choose arbitrary volumes for the threefolds at
the hidden and observable sector, if we want to preserve sensible values for
the Newton's constant and the $E_8$ gauge coupling, \witten.

Using the Liouville's measure, we can estimate the volume of the Calabi-Yau threefold at
the point $\omega_{i}\in {\cal K}(X)$ as
\foot{Being rigorous, we should work with the dimensionfull measure 
$\big(\alpha^{\prime}\omega\big)^3$,
although this will be irrelevant for our purposes because 
$\alpha^{\prime}$ factorizes out in the formulae that we use. In the small volume limit
this approximation can fail, and we should use conformal field theory to give
a more accurate estimation.
}
\eqn\voluCY{
{\rm Vol}(X)_i = {1\over 3!}\int_{X} \omega_{i}^3,
}
thus radial dilations in the K\"ahler cone
$\omega_i\mapsto N\omega_i$ with $N\in \IZ^+$, map the volume as
${\rm Vol}(X)_i \mapsto N^3 {\rm Vol}(X)_i$.

The volume of the threefolds at the boundaries of $Y$, are related through 
Witten's 
formula \witten
\eqn\wittenrel{
{\rm Vol}(X)_1 = {\rm Vol}(X)_0 + 2\pi {\rho\over\ell_P}\int_{X}\omega_{0}\wedge\Big(
c_2(V_{0})-{1\over 2}c_2(TX)
\Big) + {\cal O}(\rho^2),
}
with $\ell_P$ the eleven dimensional Planck length and $\rho$ the length of the M-theory interval.
This formula \wittenrel\ holds at first order in $\rho$, which is the limit where we work, as in \calibration.
A more accurate relation between the volumes of the CYs at the boundaries, 
taking into account the non-linear corrections in $\rho$, was derived in
\curiokrausea\ and \curiokrauseb.
The Newton's constant in the effective
supergravity theory on the observable $\IR^4$ of $Y$, goes as
\eqn\newton{
G_N \sim {\ell_P^9\over \rho {\rm Vol}(X)_0},
}  
and the $E_8$ gauge coupling as
\eqn\alphaGUT{
\alpha_{GUT}\sim {\ell_{P}^6\over {\rm Vol}(X)_0}.
}
Witten observed in \witten, that in order to find realistic values for 
these physical quantities, the volume of the threefold in the visible sector
has to be very large. As the integral in the right hand side
of \wittenrel\ is negative due to Chern-Weil theory, and the identity
\eqn\idechern{
\int_X {\rm Tr}\big( F^2\big)\wedge \omega = -\int_X \vert F\vert^2 \omega^3,
}
he deduced that sensible values for $G_N$ and $\alpha_{GUT}$ 
are only possible for 
very small ${\rm Vol}(X)_1$.  

\noindent{$ $}

\noindent{\it Summarizing:} 
Let ${\cal K}^s_0(X)$ and ${\cal K}^s_1(X)$ be the set of K\"ahler classes
that make stable $V_0\to X$ and $V_1\to X$, respectively. 
Physically interesting vacua should be located in rays of the K\"ahler cone 
lying in the intersection
${\cal K}^{s}_0(X)\cap {\cal K}^{s}_1(X)\subset {\cal K}(X)$, 
such that the relative dilating factor $\omega_0/\omega_1$ is very large, 
and the Witten's correlation
\eqn\wittencorrel{
{1\over 3!}\int_X \omega_1^3 \sim {1\over 3!}\int_X \omega_0^3 + 
2\pi {\rho\over\ell_P}\int_{X}\omega_{0}\wedge\Big(
c_2(V_{0})-{1\over 2}c_2(TX)
\Big) 
}
is satisfied.

\rmk1 Although the study of distributions of vacua 
for these models is not as developed as for Calabi-Yau
compactifications of the type $\II$ string theory,
the presence of vacua in these regions of the K\"ahler moduli space
should be statistically favorable along the lines of \douglas, once
the vector bundle, dilaton and complex moduli are stabilized.

\noindent{$ $}

\noindent{\rm We} have shown how to identify these regions explicitly. 
In the rest of the paper we determine them 
for the recently proposed Heterotic Standard Model.

\newsec{The Elliptic Calabi-Yau and its K\"ahler Cone}

First, we briefly recall the construction of the Calabi-Yau threefold used the Heterotic Standard 
Model, following the reference \ellipticCY. Let $\tilde{X}$ be the fiber product over $\IP^1$ of two rational
elliptic surfaces $\tilde{X}=B_{1}\times_{\IP^1}B_{2}$, as in the diagram: 
\eqn\diagramA{
\matrix{
&  & \tilde{X} &  &  \cr 
& \pi_{1}\swarrow &  & \searrow \pi_{2} &  \cr 
& B_{1} & \downarrow \pi & B_{2} &  \cr 
& \beta _{1}\searrow &  & \swarrow \beta _{2} &  \cr 
&  & \IP^{1} &  & \cr
}}
This kind of Calabi-Yau threefolds were already studied by C. Schoen in \S. The geometry
of $\tilde{X}$, is basically encoded in the geometry of the rational elliptic surfaces
$B_{1}$ and $B_{2}$. Due to the phenomenological interest in finding 
threefolds which admit certain Wilson lines\foot{I.e. flat line bundles with non-trivial holonomy.}, 
the aim of \ellipticCY\ was to look for threefolds $\tilde{X}$ such
that $\IZ_{3}\times \IZ_{3}\subseteq {\rm Aut}(\tilde{X})$.
This search was achieved thanks to the existence of certain elliptic surfaces that admit
an action of $\IZ_{3}\times \IZ_{3}$ which can be characterized explicitly through
a proper understanding of the Mordell-Weil group of $B$. 

Following the Kodaira's classification of singular fibers, 
our elliptic surfaces $B_{1}$ and $B_{2}$ are characterized by
three $I_{1}$ and three $I_{3}$ singular fibers. 
Such rational elliptic surfaces are described by
one-dimensional families, that allow us
to build fiber products $\tilde{X}$, corresponding to smooth Calabi-Yau threefolds. Furthermore, 
$\tilde{X}$ admits a free action of $G=\IZ_{3}\times \IZ_{3}$ and
the quotient $X=\tilde{X}/G$ is also a smooth Calabi-Yau threefold
with fundamental group $\pi_{1}(X)=\IZ_{3}\times \IZ_{3}$. 

The threefold used in the description of the Heterotic Standard Model is
$X=\tilde{X}/G$, although we will work with $G$-equivariant objects on $\tilde{X}$.
In the rest of this section we describe the $G$-invariant homology rings of 
$B$ and $\tilde{X}$, and their corresponding $G$-invariant K\"ahler cones 
(i.e. their ample cones, or spaces of polarizations).

For the homology of a surface $B$, we choose as set of generators: 
the $0$-section $\sigma$, the generic fiber $F$,
the 6 irreducible components of the three $I_{3}$ singular fibers
that do not intersect the $0$-section 
$\Theta_{1,1}, \Theta_{1,2}, \ldots \Theta_{3,1}, \Theta_{3,2}$ 
and the two sections generating the free part of the Mordell-Weil 
group\foot{See Appendix A, for a complete description of the Mordell-Weil group of
the elliptic surface.} $\xi$ and $\alpha_{B}\xi$. These generators are a basis for 
$H_{2}(B, \IZ)\otimes \IQ$, but adding the torsion generator of the 
Mordell-Weil group
\eqn\torsioneta{
\eta = \sigma + F -{2\over 3}\Big( \Theta_{1,1} + \Theta_{2,1} + \Theta_{3,1} \Big)
-{1\over 3}\Big( \Theta_{1,2} + \Theta_{2,2} + \Theta_{3,2} \Big),
}
we generate all $H_{2}(B, \IZ)$.

The intersection matrix of the homology generators is as follows:

\eqn\tableone{
\pmatrix{
    \sigma \cr F \cr
    \Theta_{1,1} \cr \Theta_{2,1} \cr \Theta_{3,1} \cr
    \Theta_{1,2} \cr \Theta_{2,2} \cr \Theta_{3,2} \cr
    \xi \cr \alpha_B \xi \cr \eta 
    \cr 
}^{T}\cdot
\pmatrix{
    \sigma \cr F \cr
    \Theta_{1,1} \cr \Theta_{2,1} \cr \Theta_{3,1} \cr
    \Theta_{1,2} \cr \Theta_{2,2} \cr \Theta_{3,2} \cr
    \xi \cr \alpha_B \xi \cr \eta 
    \cr 
}
=
\pmatrix{
    -1 & 1 & 
    0 & 0 & 0 & 0 & 0 & 0 & 
    0 & 0 & 0 
    \cr
    1 & 0 & 
    0 & 0 & 0 & 0 & 0 & 0 & 
    1 & 1 & 1 
    \cr
    0 & 0 & 
    -2 & 0 & 0 & 1 & 0 & 0 & 
    0 & 0 & 1 
    \cr
    0 & 0 & 
    0 & -2 & 0 & 0 & 1 & 0 & 
    1 & 0 & 1 
    \cr
    0 & 0 & 
    0 & 0 & -2 & 0 & 0 & 1 & 
    0 & 1 & 1 
    \cr
    0 & 0 & 
    1 & 0 & 0 & -2 & 0 & 0 & 
    0 & 1 & 0 
    \cr
    0 & 0 & 
    0 & 1 & 0 & 0 & -2 & 0 & 
    0 & 0 & 0 
    \cr
    0 & 0 & 
    0 & 0 & 1 & 0 & 0 & -2 & 
    1 & 0 & 0 
    \cr
    0 & 1 & 
    0 & 1 & 0 & 0 & 0 & 1 & 
    -1 & 1 & 0 
    \cr
    0 & 1 & 
    0 & 0 & 1 & 1 & 0 & 0 & 
    1 & -1 & 0 
    \cr
    0 & 1 & 
    1 & 1 & 1 & 0 & 0 & 0 & 
    0 & 0 & -1 \cr
}}

\noindent{$ $}

\noindent{\rm The} invariant homology under the action of $G=\IZ_{3}\times \IZ_{3}$, 
is generated by
\eqn\invhom{
H_{2}(B,\IZ)^{G}={\rm span}_{\IZ}\Big\{F,\, t= -\sigma + \Theta_{2,1}
+ \Theta_{3,1} + \Theta_{3,2} + 2\xi + \alpha_{B}\xi +\eta -F  
\Big\},
}
where $t$ can be also expressed as the homological sum of three sections, i.e.
$t=\xi + \alpha_{B}\xi + \eta\boxplus\xi$.

\noindent{The} cohomology ring of $X$, can be expressed as
\eqn\coho{
H^{\ast}(X,\,\IQ)=
H^{\ast}(\tilde{X},\,\IQ)^{G}
}
using the $G$-invariant cohomology of $\tilde{X}$.
Hence
\eqn\homology{
H^{2}(\tilde{X},\,\IQ)^{G}=\Bigg( {
H^{2}(B_{1},\, \IQ)\oplus H^{2}(B_{2},\, \IQ)\over
H^{2}(\IP^1,\, \IQ)
}\Bigg)^{G}={
H^{2}(B_{1},\, \IQ)^{G}\oplus H^{2}(B_{2},\, \IQ)^{G}\over
H^{2}(\IP^1,\, \IQ)
},
}
that due to \invhom, is the same as
\eqn\basiscohomtwo{
H^{2}(X,\,\IZ)=H^{2}(\tilde{X},\,\IZ)^{G}=
{\rm span}_{\IZ}\Big\{\tau_{1}= \pi^{*}_1 (t_{1}) ,\, \tau_{2}= \pi^{*}_2 (t_{2})  
,\, \phi =  \pi^{*}_1 (F_1) =  \pi^{*}_2 (F_1)
\Big\},
}
where $t_{1}$ and $t_{2}$ (respectively, $F_1$ and $F_2$) 
are the $t$-classes (respectively, $F$-classes) defined in \invhom, 
corresponding to each surface $B_{1}$ and $B_{2}$. Using
Poincar\'e
duality, we know that $H^{4}(X,\,\IQ)$ is isomorphic
to $H^{2}(X,\,\IQ)$, also
$H^{1}(X,\,\IZ)\simeq \pi_{1}(X)=\IZ_{3}\times \IZ_{3}$
because the Hurewicz theorem, thus
$H^{1}(X,\,\IQ)=H^{1}(X,\,\IZ)\otimes \IQ=0$.

\noindent{The} ring $H^{\ast}(\tilde{X},\,\IQ)^{G}$ generated through the cup
product of the generators \basiscohomtwo, is homomorphic to
\eqn\ringcoh{
H^{\ast}(\tilde{X},\,\IQ)^{G}=\IQ[\tau_{1},\tau_{2},\phi ]/\langle
\phi^2,\, \phi\tau_{1}=3\tau_{1}^{2},\, 
\phi\tau_{2}=3\tau_{2}^{2}
\rangle ,
}
with the top cohomology element 
being $\tau^{2}_{1}\tau_{2}=\tau_{1}\tau_{2}^{2}=3 \{ {\rm pt.} \} .$

\subsec{The Ample Cone of the Elliptic Surface.}

As first step to determine the K\"ahler cone on the threefold, we build
the $G$-invariant ample cone of the rational elliptic surface 
through the Nakai's criterion. The set of ample classes is by definition the
integral cohomology part of the K\"ahler moduli.

Using the Looijenga's classification of the effective curves in a rational
elliptic surface \effectiveclasses, we know that the 
cone of effective classes in $H_{2}(B,\,\IZ)$ is generated by the following classes
$e\in H_{2}(B,\,\IZ)$: 

$1)$ The {\it exceptional curves} $e^2:=-1$, i.e. every section of the elliptic fibration.

$2)$ The {\it nodal curves} $e^2:=-2$, i.e. the irreducible components of the singular fibers.

$3)$ The {\it positive classes}, i.e. the classes that live in 
the ``future'' side of the cone of $e^2>0$.

Nakai's criterion for surfaces says that a class
$s$ is ample if and only if $s\cdot s>0$ and 
$e\cdot s>0$ for every effective
curve $e$. We will apply this criterion to the invariant
classes $s=aF+bt$.

\noindent{$ $}

\noindent{$\bullet$} Intersection of $s$ with the {\it exceptional curves}. Although 
there is an infinite amount of {\it exceptional curves} or sections in the elliptic
surface, we can characterize them completely thanks to our understanding of the
Mordell-Weil group.

As it is explained in the Appendix A, the representation of the 
Mordell-Weil group $E(K)\simeq \IZ\oplus \IZ\oplus \IZ_{3}$ 
in ${\rm End}(H_{2}(B,\,\IZ))$, has as generators:
$(t_{\xi})_{*}$, $(t_{\alpha_{B}\xi})_{*}$ and $(t_{\eta})_{*}$. Thus, the homology of
an arbitrary section can be expressed as
\eqn\homsec{
\big[\boxplus x\xi\boxplus y\alpha_{B}\xi\boxplus z\eta
\big] =
(t_{\xi})_{*}^{x}(t_{\alpha_{B}\xi})_{*}^{y}(t_{\eta})_{*}^{z}\sigma
}
where $\boxplus x\xi$ (respectively $\boxplus y\alpha_{B}\xi$ and $\boxplus z\eta$) 
means 
$\boxplus x\xi=\underbrace{\xi\boxplus\xi\boxplus\dots \boxplus\xi}_{x}$.

Finding the Jordan canonical forms associated to
$(t_{\xi})_{*}$, $(t_{\alpha_{B}\xi})_{*}$ and $(t_{\eta})_{*}$,
allows us to expand \homsec, explicitely. We exhibit the list of homology
classes associated to the sections in the Appendix A. Hence, the intersections of 
the {\it exceptional curves} with the generators of the invariant homology are
\eqn\interone{
F\cdot \big[\boxplus x\xi\boxplus y\alpha_{B}\xi\boxplus z\eta \big]=1
}
and 
\eqn\intertwo{
t\cdot \big[\boxplus x\xi\boxplus y\alpha_{B}\xi\boxplus z\eta \big]=x^{2}+y^{2} - xy - x.
}
It is easy to check that $x^{2}+y^{2} - xy - x$, as a function 
$\IZ\oplus \IZ\to \IZ$, is non-negative and becomes zero for
$(x=0,\, y=0)$, $(x=1,\, y=0)$ and $(x=1,\, y=1)$. Therefore a $G$-invariant
ample class $s=aF+bt$, has to verify
\eqn\firam{
s\cdot \big[ \boxplus 0\xi\boxplus 0\alpha_{B}\xi\boxplus z\eta \big]=
a>0,
}
and
\eqn\secam{
s\cdot \big[ \boxplus\infty\xi\boxplus\infty\alpha_{B}\xi\boxplus z\eta \big]=
a+\infty b>0,\, \Rightarrow\, b>0.
}

\noindent{$ $}

\noindent{$\bullet$} Intersection of $s$ with the {\it nodal curves}. The nodal curves
are identified with the irreducible components $\Theta_{i,j}$ of the singular fibers, 
thus their intersections with the invariant class $s=aF+bt$ give us 
\eqn\nodalint{
s\cdot \Theta_{i,j} = b > 0.
}
Identical result to the inequality \secam, derived above. 

\noindent{$ $}

\noindent{$\bullet$} Intersection of $s$ with the {\it positive classes}. Let ${\cal K}^{+}(B)$ be the
cone of positive classes in $B$, i.e. ${\cal K}^{+}(B)=\{ e\in  H_{2}(B,\IZ)\vert\, e\cdot e>0 \}$.
As ${\cal K}^{+}(B)$ is a convex set and we have to take intersections
of elements in ${\cal K}^{+}(B)$ with invariant classes in $H_{2}(B,\IZ)^{G}$,
only the intersection ${\cal K}^{+}(B)\cap H_{2}(B,\IZ)^{G}$ matters.
From the intersection matrix of the homology generators, we know that the intersection matrix
of the invariant homology $H_{2}(B,\IZ)^{G}$ is
\eqn\tabletwo{
\pmatrix{
F\cr
t\cr
}^{T}\cdot
\pmatrix{
F\cr
t\cr
}
=\pmatrix{
0&3\cr
3&1\cr
}
} 
hence, we find
\eqn\invariantposi{
{\cal K}^{+}(B)\cap H_{2}(B,\IZ)^{G}:=\big\{ 
e=xF+yt\vert\, 6xy + y^2 > 0
\big\},
} 
being the edges of such ``future'' cone $F$ and $6t-F$.
Furthermore, their intersections with our ample candidate $s=aF+bt$, 
give us the conditions 
\eqn\interpos{\eqalign{
s\cdot F=(aF+bt)\cdot F = 3b > 0 \cr
s\cdot (6t-F)=18a+6b-3b=18a+3b > 0
}}
that do not constrain the inequalities \firam, and \secam.

Finally, as the cone generated by $F$ and $t$ is within 
${\cal K}^{+}(B)\cap H_{2}(B,\IZ)^{G}$, the last
Nakai's condition $s\cdot s>0$ or positivity
of the Liouville's measure, is verified. 
Therefore, the $G$-invariant ample cone associated to the elliptic surface
$B$ is simply 
\eqn\amplecone{
{\cal K}(B)^{G}={\rm span}_{\IZ^{+}}\big\{
F,\, t
\big\}. 
}

\subsec{Ampleness in the Threefold.}

Once we have characterized the $G$-invariant ample cone on the rational surface, 
we can construct $G$-invariant ample classes on the threefold $\tilde{X}$ as product of
ample classes on the surfaces $B_{1}$ and $B_{2}$. 
In fact, the following proposition shows that
the amples classes on $\tilde{X}$ constructed in this way 
determine explicitely its $G$-invariant ample cone ${\cal K}(\tilde{X})^{G}={\cal K}(X)$.

\prop{3.1}
{\it
The $G$-invariant ample cone of $\tilde{X}$ is}
\eqn\acxtile{
{\cal K}(\tilde{X})^{G} = {\rm span}_{\IZ^{+}}\big\{
\tau_1, \tau_2, \phi
\big\}. 
}

\noindent{\it Proof.} 
If $L_i$ is an ample class in $B_i$, then $\pi_1^* L_1\otimes \pi_2^* L_2$
is an ample class in $\tilde X$, hence ${\cal K}(\tilde{X})^{G}$ contains
the positive linear span of $\tau_1$, $\tau_2$ and $\phi$.

To show the opposite inclusion, we apply Nakai's criterion to some 
effective classes. Let $H=a\tau_1+b\tau_2+c\phi$ be an ample class.
If $C_1$ be the class of a fiber of $\pi_1$,
$$
0< H\cdot C_1= 0 a + 3 b + 0 c = 3b
$$
Analogously, if $C_2$ is the class of a fiber of $\pi_2$, we obtain $a>0$.
Let $i:B_1\times_{\IP^1}B_2\to B_1\times B_2$. Let $C$ be the class
of $\sigma_1\times_{\IP^1}\sigma_2$, let $c_1$, $c_2$ be two integers
with $c=c_1+c_2$, and denote $[B_i]$ (respectively, $[{\rm pt}]$) 
the class of $B_i$ in $H^0(B_i,\IZ)$ (respectively, 
of a point in $H^4(B_i,\IZ)$).
\eqn\segundo{\eqalign{
0<H\cdot C=
i^* \big( (a t_1+c_1 f_1)\otimes[B_2] + [B_1]\otimes(bt_2+c_2f_2) \big)\cdot
i^* [\sigma_1\otimes \sigma_2]\cr
= i^* \Big(
(at_1+c_1f_1)\sigma_1 \, [{\rm pt}]\otimes\sigma_2 +
\sigma_1\otimes[{\rm pt}] \, (bt_2+c_2f_2)\sigma_2
\Big)\cr
= i^*\Big( 
c_1 \, [{\rm pt}]\otimes\sigma_2 +
c_2 \,\sigma_1\otimes[{\rm pt}]
\Big)\cr
=c_1+c_2 = c
}}

\hfill $\spadesuit$

\newsec{Slope Stability of the Vector Bundles}

The concept of (slope) stability of a vector bundle depends on the choice of a polarization
$H \in {\cal K}(X)\subset H^{2}(X,\, \IZ)$,
i.e., we say that a holomorphic vector bundle ${\cal E}\to X$ 
is stable iff
\eqn\stable{
\mu({\cal F}) < \mu({\cal E}) \, ;\quad {\rm with}\quad \mu(\cdot)={H^2\cdot {\rm det}(\cdot) \over
{\rm rank}\,(\cdot)},
}
for every reflexive subsheaf ${\cal F}\to {\cal E}$. By
${\rm det}({\cal E})$  and ${\rm det}({\cal F})$ we mean the determinant 
line bundles associated to ${\cal E}$ and ${\cal F}$. 

There is a natural bijection between vector bundles on $X$ and $G$-equivariant
vector bundles on $\tilde{X}$.
We will recall a few general remarks on $G$-invariance and
$G$-equivariance, which will be useful in the rest of this section.

Let $X$ be a complex projective variety and $G$ a complex algebraic group
acting on it.
A subvariety $X^{\prime }$ of $X$ is said invariant if $gX^{\prime
}=X^{\prime }$ for all $g$ in $G$. A divisor ${\cal D}$ is said invariant if $g{\cal D}={\cal D}$
for all $g$ in $G$. A divisor class is said invariant if for any divisor 
${\cal D}$ in the class and $g$ and in $G$, the divisor $g{\cal D}$ is linearly
equivalent to ${\cal D}$.

An equivariant structure on a vector bundle $E$ on $X$ is a lifting, by
linear maps $E(x)\longrightarrow E(gx)$ (for all $g\in G$) between fibers,
of the action of $G$ on $X$. We will widely use this notion, and sometimes
also the notion of equivariant coherent sheaf (will talk about some
equivariant ideal sheaf) so it is convenient to generalize it defining an
equivariant structure on a coherent sheaf $F$ on $X$ as a family of
isomorphisms $\varphi_{g}^{F}:$ $F$ $\cong $ $g^{\ast }F$ , for each 
$g\in G$, so that $\varphi_{g^{\prime }g}^{F}=\varphi _{g}^{F}\varphi _{g^{\prime}}^{F}$. Equivariant morphisms 
\eqn\morphdef{f:F\longrightarrow F^{\prime},} 
between equivariant sheaves are
those such that 
\eqn\equshea{
\matrix{
 & \varphi_{g}^{F} & \cr
F & \longrightarrow  & g^{\ast }F \cr
f\downarrow &  & g^{\ast }f\downarrow \cr
F^{\prime } & \longrightarrow & 
g^{\ast }F^{\prime } \cr
 & \varphi_{g}^{F^{\prime}} & \cr
}}
for all $g\in G$. 
\bigskip

If two vector bundles have an equivariant structure, obviously their tensor
products inherit an equivariant structure. If a vector bundle $E$ has an
equivariant structure, all of its exterior powers, and in particular its
determinant line bundle, ${\rm det}\,( E)$, inherit an equivariant
structure, and also its dual $E^{\ast }$ (pointwise, take the inverse of the
transposed action). The trivial bundle $L= X \times \IC$, or ${\cal O}_{X}$ 
as associated sheaf, admits a trivial equivariant structure.

A vector bundle with equivariant structure is always {\it invariant},
which means, by definition, that $g^{\ast }E$ is isomorphic to $E$ for any 
$g$ in $G$ . In the case $E$ is a vector bundle $L$ of rank 1,
this definition means that both $g^{\ast }L$ and $L$ define the same point
of ${\rm Pic} (X)$, i.e. that the point corresponding to $E$ in ${\rm Pic} (X)$ 
is fixed by the action of the group, or still, in terms of associated divisors,
that the corresponding divisor {\it class} is invariant.

A vector subbundle $E^{\prime }\subset E$ of an equivariantly structured
bundle $E$ is called an equivariant vector subbundle when 
$g(E^{\prime}(x))\subset E^{\prime}(x)$ for all $x$ in $X$ and $g$ in $G$. 
This is equivalent to say that, for all $g\in G$, the isomorphism 
$E \cong g^{\ast }E$ given by the equivariant structure, applies $E^{\prime }$ into 
$g^{\ast }E^{\prime }$, so this notion still has a meaning when $E^{\prime}$
is just a coherent subsheaf. An equivariant coherent subsheaf $E^{\prime}$
obviously inherits an structure of equivariant coherent sheaf, as well as
its quotient $E^{\prime \prime }=E/E^{\prime }$, and we just say that the
extension 
\eqn\exteqone{
0\longrightarrow E^{\prime }\longrightarrow E\longrightarrow E^{\prime
\prime }\longrightarrow 0,
}
is equivariant.

An equivariant vector bundle is said equivariantly stable if all its
equivariant coherent sheaves (enough to check with reflexive) have smaller
slope. A section $s$ of equivariantly structured $E$ is called equivariant when,
for all $x$ in $X$ and $g$ in $G$, it is $g(s(x))=s(gx)$ . When viewing the
section, as usual, as a subbundle ${\cal O}_X\rightarrow E$, this
amounts to say that the the subbundle is equivariant and the inherited
equivariant structure on the trivial bundle is the trivial equivariant
structure. Clearly, the vanishing locus $V(s)$ of an equivariant section is
invariant. If the vector bundle $E$ is a line bundle $L$ of rank one, and $s$
is a meromorphic equivariant section of it, i.e. equivariant section defined
on a Zariski open set, the divisor it defines is an invariant divisor (not
only a divisor of invariant class). We say $L$ is equivariantly effective if
it has a nonzero equivariant global (i.e. holomorphic)\ section.

In a surface $X$, a line bundle $L={\cal O}_X(D)$ is
equivariantly ample when is equivariant and has positive
selfinterseccion, and its itersection number with all equivariantly
effective equivariant line bundles is positive. Therefore, 
ample and equivariant implies equivariantly ample.

\subsec{Conditions on the Effective Divisors}

This is an analysis previous to the solution of both problems. We show now
that if there exists an efective divisor in the invariant class 
${\cal O}_{B}(at+bF)$ on
the elliptic surface, then $a\geq -3b.$
We start with the following:

\noindent{$ $}
\rmk2 Denote $a^{\prime }$ the defect quotient 
\eqn\rema{
a^{\prime }=\left[ {a\over 3}\right],
}
Recall that $t$ is the homology sum of three sections,
namely $\xi$, $\alpha_B\xi$ and $\eta\boxplus\xi$, which
we denote, respectively, $s_1$, $s_2$ and $s_3$.
The $3a$ summands in 
\eqn\summand{
at=as_{1}+as_{2}+as_{3}
}
can be ordered
\eqn\sumsum{
at=s_{1}^{\prime }+...+s_{3a}^{\prime },
}
so to fullfill the following three conditions:

\noindent{$\bullet$} For all index $i$ such that $s_{i}^{\prime }=s_{1}$ 
\eqn\bullone{
\sharp \{s_{j}^{\prime }\mid j\leq i{\rm \, and \, }s_{j}^{\prime}
=s_{2}\}-\{j\mid j\leq i{\rm \, and \, }s_{j}^{\prime }=s_{1}\}\leq
a^{\prime }.
}

\noindent{$\bullet$} For all index $i$ such that $s_{i}^{\prime }=s_{3}$, 
\eqn\bulltwo{
\sharp \{s_{j}^{\prime }\mid j\leq i {\rm \, and \, }s_{j}^{\prime
}=s_{2}\}-\{j\mid j\leq i {\rm \, and \, }s_{j}^{\prime }=s_{3}\}\leq
a^{\prime }.
}

\noindent{$\bullet$} For all index $i$ such that $s_{i}^{\prime }=s_{2}$, 
\eqn\bullthree{
\sharp \{s_{j}^{\prime }\mid j\leq i {\rm \, and \, }s_{j}^{\prime }=s_{1}%
{\rm\, or \, }s_{3}\}-\{j\mid j\leq i {\rm \, and \, }s_{j}^{\prime
}=s_{2}\}\leq a^{\prime }.
}

Indeed, the following ordering of the $3a$ summands satisfies the three
conditions: take its first $3a^{\prime }$ summands to be 
\eqn\sumofsections{
(s_{1}+s_{2}+s_{3})+\dots +(s_{1}+s_{2}+s_{3}).
}
Next, add summands of the alternating form
\eqn\sumsecond{
(s_{1}+s_{2})+(s_{3}+s_{2})+(s_{1}+s_{2})+(s_{3}+s_{2})+\ldots
}
(so $s$ has already ocurred $a$ times) and add finally summands $s_{1},s_{3}$, 
in no matter which order, until completing $a$ ocurrences of each.

\noindent{$ $}

The consequence of this remark is the following

\lem1 : For any direct factor ${\cal O}_{\IP^1}(l)$
ocurring in the splitting of $\beta _{\ast }{\cal O}_{B}(at)$ it is 
$l\leq a^{\prime}:=\left[ {a\over 3} \right]$, i.e. 
$h^{0}(\beta_{\ast }{\cal O}_{B}(at)(-a^{\prime }-1))=0$.

\noindent{\it Proof.} Recall $\beta_{\ast} {\cal O}_{B}={\cal O}_{\IP^1}$.
Order the $3a$ summands in 
\eqn\sumfour{
at=s_{1}^{\prime }+\ldots +s_{3a}^{\prime },
}
as in the former remark. For some index $1\leq i<3a$, assume it is already
proved that 
\eqn\alreadyproved{
h^{0}(\beta_{\ast }{\cal O}_{B}(s_{1}^{\prime }+
\ldots +s_{i-1}^{\prime
})(-a^{\prime }-1))=0.
}
It is then enough to prove that
\eqn\sumofsec{
h^{0}(\beta_{\ast}{\cal O}_{B}(s_{1}^{\prime }+\ldots +s_{i}^{\prime
})(-a^{\prime }-1))=0.
}
From 
\eqn\severalaarrow{
0\longrightarrow {\cal O}_{B}(s_{1}^{\prime }+\ldots +s_{i-1}^{\prime
})\longrightarrow {\cal O}_{B}(s_{1}^{\prime }+\ldots +s_{i}^{\prime
})\longrightarrow {\cal O}_{s_{1}}(s_{1}^{\prime }+\ldots +s_{i}^{\prime
})\longrightarrow 0,
}
we obtain 
\eqn\ssss{
0\longrightarrow \beta _{\ast}{\cal O}_{B}(s_{1}^{\prime
}+\ldots +s_{i-1}^{\prime })\longrightarrow \beta _{\ast} {\cal O}_{B}
(s_{1}^{\prime }+\ldots +s_{i}^{\prime })\longrightarrow 
{\cal O}_{\IP^1}((s_{1}^{\prime }+\ldots 
+s_{i}^{\prime })s_{1})\longrightarrow 0.
}
Assume first that $s_{i}^{\prime }=s_{1}$. Recalling that $s_{1}^{2}=-1$, 
$s_{1}s_{3}=0,s_{1}s_{2}=1$, we have
\eqn\pppp{
(s_{1}^{\prime }+\ldots +s_{i}^{\prime })s_{1}=\sharp \{j\mid j\leq i{\rm \, and \,}
s_{j}^{\prime }=s_{2}\}-\sharp \{j\mid j\leq i {\rm \, and \,}s_{j}^{\prime}
=s_{1}\}\leq a^{\prime },
}
proving, by consulting the former exact sequence, the wanted vanishing. The
vanishing is analogously proved in the case $s_{i}^{\prime }=s_{3}$.

Assume now that $s_{i}^{\prime }=s_{2}$. Recalling that $s_{2}^{2}=-1$, 
$s_{2}s_{1}=s_{2}s_{3}=1$, we have
\eqn\gggggg{
(s_{1}^{\prime}+\ldots +s_{i}^{\prime })s_{2}=\sharp \{j\mid j\leq i{\rm \, and \,}
s_{j}^{\prime }=s_{1} {\rm \, or \,}s_{3}\}-\sharp \{j\mid j\leq i\ {\rm \, and \,}
s_{j}^{\prime }=s_{2}\}\leq a^{\prime }
}
thus proving also the wanted vanishing \hfill $\spadesuit$. 

\bigskip

\noindent{\bf Corollary.} If 
\eqn\wewe{
H^{0}(B, {\cal O}_{B}(at+bF))\neq 0,
}
then  $a\geq -3b$.

\noindent{\it Proof.} We assume 
\eqn\hjhjhj{
0\neq H^{0}(B,{\cal O}_{B}(at+bF))=H^{0}(\IP^1 ,\beta_{\ast }
{\cal O}_{B}(at+bF))=H^{0}(\IP^1,{\cal O}_{\IP^1}(b)
\otimes \beta_{\ast}{\cal O}_{B}(at)),
}
where $\beta_{\ast }{\cal O}_{B}(at)$ is a direct sum of factors 
${\cal O}_{\IP^1}(l)$ with $l\leq a/3$, by the lemma.
Therefore, for some of these factors, we obtain 
\eqn\qwasqw{
0\leq b+l\leq b+{a\over 3}
}

\hfill $\spadesuit$.

\lem2 : a) If 
$
H^{0}(\tilde{X}, {\cal O}_{\tilde{X}}(a_1\tau_1+a_2\tau_2+b\phi))\neq 0,
$
then $a_{1},a_{2}\geq 0$ and $b\geq -{1\over 3}(a_{1}+a_{2})$.

b) If $H^0(\tilde{X},{\cal I}_{\Theta}(a_1\tau_1+b\phi))\neq 0$,
then $a_1\geq 0$, $b\geq -{1\over 3}a_1+3$.

\noindent{\it Proof.} a) If $a_{i}$ were negative, then the restriction of this
section to any elliptic fibre $E_{i}$ of $\pi_{2}$ would be
\eqn\sedfr{
{\cal O}_{E_{i}}\longrightarrow {\cal O}%
_{E_{i}}(a_{1}(p_{1}+p_{2}+p_{3})),
}
and this is impossible.
\noindent{\rm On} the other hand, 
\eqn\mkmkwmk{
\matrix{
H^{0}\big( {\cal O}_{\tilde{X} }(a_1\tau_1+a_2\tau_2+b\phi) \big) & = & 
H^{0}\big( {\cal O}_{B_{1}}(a_{1}t_1+bF_1)\otimes \pi_{1\ast }\pi_{2}^{\ast}
{\cal O}_{B_{2}}(a_{2}t_2)\big) \cr
 & = & H^{0}({\cal O}_{B_{1}}(a_{1}t_1+bF_1)\otimes 
\beta_{1}^{\ast}\beta_{2\ast
}{\cal O}_{B_{2}}(a_{2}t_2)) \cr
 & = &H^{0}(\beta_{1\ast }{\cal O}_{B_{1}}(a_{1}t_1)\otimes 
 {\cal O}_{\IP^1}( b)\otimes \beta_{2\ast }{\cal O}_{B_{2}}(a_{2}t_2)) \cr
 & = & H^{0}(\bigoplus_{i}{\cal O}_{\IP^1}(l_{1i})
 \otimes {\cal O}_{\IP^1}(b)\otimes
\bigoplus_{j}{\cal O}_{\IP^1}(l_{2j})).\cr
}}
In these sums $l_{1i}\leq a_{1}^{\prime }:=\left[ {a_{1}\over 3}\right] $
and $l_{2j}\leq a_2 ^{\prime}:=\left[ {a_{2}\over 3}\right]$, 
because of the former lemma, so if this is nonzero, 
then for some direct factors 
${\cal O}_{\IP^1}(l_{1})$ and ${\cal O}_{\IP^1}
(l_{2})$ appearing in the decomposition it is
\eqn\dfere{
0\leq l_{1}+b+l_{2}\leq {a_{1}\over 3}+b+{a_{2}\over 3}
}

b) Remark that 
\eqn\aliii{
\pi^{}_{1*}\pi_2^* {\cal I}_\Theta = \beta^*_1\beta_{2*}{\cal I}_\Theta
= \beta^*_1 {\cal O}_{\IP^1}(-3) = {\cal O}_{\tilde{X}}(-3\phi),
}
since $\beta_{2*}{\cal I}_\Theta ={\cal O}_{\IP^1}(-p_1-p_2-p_3)
\cong  {\cal O}_{\IP^1}(-3)$ (because $Z$ lies in the fibers of $\beta_2$
at three different points $p_1$, $p_2$, $p_3\in \IP^1$).
Therefore
\eqn\bliii{
0 \neq H^0(\tilde{X},{\cal I}_{\Theta}(a_1\tau_1+b\phi)) =
H^0(B_1, \pi^{}_{1*}\pi_2^* {\cal I}_\Theta \otimes{\cal O}_{B_1}(a_1\tau_1+b\phi)) =
H^0(B_1,  {\cal O}_{B_1}(a_1\tau_1+(b-3)\phi))
}
and we conclude using the previous Corollary.

\hfill $\spadesuit$.

\subsec{The Hidden bundle.}

Let ${\cal H}$ be a rank-$2$ subbundle of the vector bundle
${\cal E}_{\rm h}\to X$, adjoint representation of the hidden $E_{8}$ gauge group, 
defined through the short exact sequence
\eqn\hiddenseq{
0\longrightarrow
{\cal O}_{\tilde{X}}(2\tau_{1}+\tau_{2}-\phi)
\longrightarrow
{\cal H}
\longrightarrow
{\cal O}_{\tilde{X}}(-2\tau_{1}-\tau_{2}+\phi)
\longrightarrow
0.
}
By construction of the extension, the determinant line bundle associated to ${\cal H}$ is
trivial, thus the slope of the rank-$2$ vector bundle is $\mu({\cal H})=0$.
On the other hand, ${\cal O}_{\tilde{X}}(2\tau_{1}+\tau_{2}-\phi)$ admits a morphism
to ${\cal H}$ as it is shown in the diagram \hiddenseq, therefore given a polarization
$H={\cal O}_{\tilde{X}}(x\tau_{1}+y\tau_{2}+z\phi)$ with $x,\, y,\, z\in \IZ^{+}$, we have
\eqn\slopeline{
\mu \big( {\cal O}_{\tilde{X}}(2\tau_{1}+\tau_{2}-\phi)\big) =
H^{2}\cdot {\cal O}_{\tilde{X}}(2\tau_{1}+\tau_{2}-\phi)=
3(x^2 + 2y^2 +6xz + 12yz)>0.
}
that is positive for all $H\in {\cal K}(X)$ thus
$\mu \big( {\cal O}_{\tilde{X}}(2\tau_{1}+\tau_{2}-\phi)\big)>\mu ({\cal H})$,
what means that ${\cal O}_{\tilde{X}}(2\tau_{1}+\tau_{2}-\phi)$ is
a destabilizing line bundle for ${\cal H}$.
As ${\cal H}$ is not stable, we cannot integrate the hermitian Yang-Mills
equations in order to construct an $SU(2)$-instanton on ${\cal H}$. 
We must substitute ${\cal H}$ in order to find a sensible vacuum
for the heterotic string.

\subsec{The Visible bundle.}

Here, we recall the construction of the visible bundle, \upennbundle. First
it is defined 
an equivariant rank 2 vector bundle $V_{2}$
on $B$ of trivial determinant given as nontrivial extension 
\eqn\nontrivial{
0\longrightarrow {\cal O}_{B}(-2F)\longrightarrow V_{2}\longrightarrow 
{\cal I}_{Z}(2F)\longrightarrow 0,
}
with $Z$ the scheme of 9 points,
together with an equivariant structure on $V_{2}$ so that this extension
is equivariant
\eqn\extetwo{
0\longrightarrow {\cal O}_{\tilde{X}}(-2\phi)\longrightarrow 
\pi_{2}^{\ast}V_{2}\longrightarrow {\cal I}_{\Theta }(2\phi)
\longrightarrow 0.
}
Here, $\Theta $ the lifting to $\tilde{X}$ of $Z$ by the second
projection.
Then, the visible rank $4$ vector bundle $V_{4}$ of trivial determinant, is defined 
through the extension
\eqn\extthre{
0\longrightarrow {\cal O}(-\tau_1+\tau_2)\oplus {\cal O}(-\tau_1+\tau_2)%
\longrightarrow V_{4}\longrightarrow V_{2}(\tau_1-\tau_2)\longrightarrow 0,
}
together with an equivariant structure making this extension equivariant,
and general among such extensions.

We will show there exists some equivariant line bundle 
${\cal O}_{\tilde{X}}(x_{1}\tau _{1}+x_{2}\tau _{2}+y\phi)$,
thus of corresponding class of
divisors $H$ being invariant, i.e. $H=x_{1}\tau _{1}+x_{2}\tau _{2}+y\phi$,
such that the integes $x,y,z$ are positive (thus ${\cal O}_{\tilde{X}}
(x_{1}\tau _{1}+x_{2}\tau _{2}+y\phi)$ 
equivariantly ample) and making the equivariant
bundle $V_{4}$ equivariantly stable.

The degree of a line bundle ${\cal O}_{\tilde{X}}(a_1\tau_1+a_2\tau_2+b\phi)$, respect to the
polarization $H$ is 
\eqn\polarizat{
\matrix{
H^{2}(a_{1}\tau _{1}+a_{2}\tau _{2}+bf)
&=
&3(x_{1}+x_{2}+6y)(a_{1}x_{2}+a_{2}x_{1})+x_{1}x_{2}(3a_{1}+3a_{2}+18b) \cr
&=&3x_2(2x_1+x_2+6y)a_1 + 3x_1(x_1+2x_2+6y)a_2 + 6x_1 x_2 b\cr
}}
Clearly, this degree function is strictly monotonous with respect to 
the obvious
partial ordering among these line bundles or triples of integers $(a_{1},a_{2},b)$.
Now we will list all possible subsheaves.

\bigskip

\noindent{\bf 1).} {\it Possible line subbundles.}

For this, we see first necessary conditions on $a_{1},a_{2},b$ for 
$\pi_{2}^{\ast }V_{2}$ to admit ${\cal O}_{\tilde{X}}(a_1\tau_1+a_2\tau_2+b\phi)$ as equivariant
line subbundle
\eqn\subdu{\matrix{
0 & \longrightarrow & {\cal O}_{\tilde{X}}(-2\phi) & \longrightarrow & 
\pi_{2}^{\ast}V_{2} & \longrightarrow & {\cal I }_{\Theta }(2\phi)
& \longrightarrow & 0 \cr 
&  &  &  & \uparrow &  &  &  &  \cr 
&  &  &  & {\cal O}_{\tilde{X}}(a_1\tau_1+a_2\tau_2+b\phi) &  &  &  & \cr 
}}

If $a_{1}\leq 0$ and $a_{2}\leq 0$ and $b\leq -2-{1\over 3}(a_{1}+a_{2})$
is not fulfilled, then the intersection of this subbundle with the one on
the left must be null, so ${\cal O}_{\tilde{X}}(a_1\tau_1+a_2\tau_2+b\phi)$ becomes an
equivariant subsheaf of the one in the right, thus giving an equivariant
nonzero section of ${\cal O}_{\tilde{X}}(-a_{1}\tau_1-a_{2}\tau_2+b\phi)$ 
vanishing at 
$\Theta$. We thus get possibilities
\eqn\possib{\matrix{
{\rm i) } & &a_{1}\leq 0 {\rm \, and \,}a_{2}\leq -1 {\rm \, and \,}
b\leq 2- {1\over 3}(a_{1}+a_{2}) \cr
{\rm  ii) }& &a_{1} \leq 0 {\rm \, and \,} a_{2}=0 {\rm \, and \,} b\leq -1-
{1\over 3}a_{1} \cr
{\rm iii) } &&a_{1}\leq 0 {\rm \, and \,} a_{2}\leq 0 {\rm \, and \, }b\leq 
-2-{1\over 3}(a_{1}+a_{2}). \cr
}}
For ii) we have used Lemma 2 b).
Let us find now necessary conditions for the existence of an equivariant
rank 1 reflexive sheaf, i.e. equivariant subbundle ${\cal O}_{\tilde{X}}
(a_1\tau_1+a_2\tau_2+b\phi)$, of $V_{4}$:
\eqn\reffre{\matrix{
0 & \longrightarrow & {\cal O}_{\tilde{X}}(-\tau_1+\tau_2)\oplus {\cal O}_{\tilde{X}}(
-\tau_1+\tau_2) & \longrightarrow & V_{4} & \longrightarrow & \pi_{2}^{\ast
}V_{2}(\tau_1-\tau_2) & \longrightarrow & 0 \cr 
&  &  &  & \uparrow &  &  &  &    \cr 
 &  &  &  & & {\kern-6.5em{\cal O}_{\tilde{X}}(a_1\tau_1+a_2\tau_2+b\phi)}   &  &  &  \cr
}}

\noindent{By} the same argument as above, combined with our former discusion
on equivariant line subbundles of $\pi^*_{2}V_{2}$, we obtain these possibilities:

\eqn\wqwq{\matrix{
{\rm i.1)} & &a_{1}\leq -1 {\rm \, and \, }a_{2}\leq 1 {\rm \, and \,}
b\leq- {1\over 3}(a_{1}+a_{2}) \cr
{\rm i.2)} & &a_{1}\leq 1 {\rm \, and \, }a_{2}\leq -2 {\rm \, and \, }
b\leq 2- {1\over 3}(a_{1}+a_{2}) \cr
{\rm i.3) }& &a_{1} \leq 1{\rm \, and \,}a_{2}=-1{\rm \, and \,}b\leq -{2\over 3}-{1\over 3}
a_{1} \cr
{\rm i.4)} & &a_{1}\leq 1{\rm \, and \,}a_{2}\leq -1{\rm \, and \, }%
b\leq -2-{1\over 3}(a_{1}+a_{2}). \cr
}}

\bigskip

\noindent{\bf 2).} {\it Possible reflexive sheaves of rank 2.}

Let us consider now an equivariant reflexive subsheaf of rank 2
\eqn\dwdw{\matrix{
0 & \longrightarrow & {\cal O}_{\tilde{X}}(-\tau_1+\tau_2)\oplus {\cal O}_{\tilde{X}}\left(
-\tau_1+\tau_2\right) & \longrightarrow & V_{4} & \longrightarrow & 
\pi_{2}^{\ast}V_{2}(\tau_1-\tau_2) & \longrightarrow & 0 \cr 
&  &  &  & \uparrow &  &  &  &  \cr 
&  &  &  & R_{2} &  &  &  & \cr
}}
having nonnegative degree. Since all of its equivariant line subbundles, as
equivariant subbundles of $V_{4}$, must have, as seen, negative degree, 
the reflexive sheaf 
$R_{2} $ is equivariantly semistable. If its intersection with the subbundle
of $V_{4}$ in its above presentation were not zero, then there would be a
nonzero equivariant morphism
\eqn\qwmilq{
R_{2}\longrightarrow {\cal O}_{\tilde{X}}(-\tau_1+\tau_2)\oplus {\cal O}_{\tilde{X}}
\left(-\tau_1+\tau_2\right),
}
between both equivariantly semistable sheaves, so that the first should have
slope not bigger than the slope of the second, i.e. $R_{2}$ should have
degree not bigger than the degree of the direct sum, which is negative
(as seen in the former step). We thus obtain an injection 
\eqn\injectione{
0\longrightarrow R_{2}\longrightarrow V_{2}(\tau_1-\tau_2)\longrightarrow
Q\longrightarrow 0,
}
between these equivariant reflexive sheaves of rank 2, thus its quotient $Q$
is a torsion sheaf. We thus obtain a nonzero equivariant morphism
\eqn\morphh{
{\cal O}_{\tilde{X}}(a_1\tau_1+a_2\tau_2+b\phi)=\bigwedge^{2}R_{2}\longrightarrow
\bigwedge^{2}V_{2}(\tau_1-\tau_2)={\cal O}_{\tilde{X}}(2\tau_1-2\tau_2).
}
Therefore, necessarily
\eqn\nede{
{\rm ii.1) \,} a_{1}\leq 2{\rm \, and \, }a_{2}\leq -2{\rm \, and\, }b\leq 
-{1\over 3}(a_{1}+a_{2}).
}
The top case 
$
a_{1}=2{\rm \, and \,}a_{2}=-2{\rm \, and \,}b=0
$,
would give a contradiction to what we want to prove, if it ocurred, as no
polarization of $\tilde{X}$ giving to ${\cal O}(-\tau_1+\tau_2)\oplus {\cal O}(-\tau_1+\tau_2)$
negative degree would give negative degree to 
${\cal O}_{\tilde{X}}(2\tau_1-2\tau_2)$, but
fortunately it does not occur. Indeed, if this were the case, then the
quotient $Q$ would be supported in codimension at least two, but this is
incompatible with the kernel $R_{2}$ of such a quotient being reflexive,
unless $Q=0$, i.e. $R_{2}\cong V_{2}(\tau_1-\tau_2)$, thus splitting the sequence
presenting $V_{4}$. This would contradict the genericity of the extension
taken in its presentation. Therefore, we get three subcases:
\eqn\neded{\matrix{
&{\rm ii.1.a) \,}& a_{1}\leq 1{\rm \, and \, }a_{2}\leq -2{\rm \, and\, }b\leq 
-{1\over 3}(a_{1}+a_{2}).\cr
&{\rm ii.1.b) \,}& a_{1}\leq 2{\rm \, and \, }a_{2}\leq -3{\rm \, and\, }b\leq 
-{1\over 3}(a_{1}+a_{2}).\cr
&{\rm ii.1.c) \,}& a_{1}\leq 2{\rm \, and \, }a_{2}\leq -2{\rm \, and\, }b\leq 
-1-{1\over 3}(a_{1}+a_{2}).\cr
}}

\bigskip

\noindent{\bf 3).} {\it Possible rank 3 equivariant reflexive sheaves.}


We can consider these equivariant subsheaves saturated, i.e. having as
quotient a rank 1 torsion free sheaf, so with a line bundle 
${\cal O}_{\tilde{X}}(a_1\tau_1+a_2\tau_2+b\phi)$ as dual. In other words, giving such a subsheaf is
equivalent to giving an equivariant line subbundle as in the diagram
\eqn\sedfer{\matrix{
0 & \longrightarrow & \pi _{2}^{\ast }V_{2}(-\tau_1+\tau_2) & \longrightarrow & 
V_{4}^{\vee } & \longrightarrow & {\cal O}_{\tilde{X}}(\tau_1-\tau_2)\oplus 
{\cal O}_{\tilde{X}}(\tau_1-\tau_2) & \longrightarrow & 0 \cr 
&  &  &  & \uparrow &  &  &  &  \cr
&  &  &  &  &{\kern-6.5em{{\cal O}_{\tilde{X}}(a_1\tau_1+a_2\tau_2+b\phi)}}  &  &  & \cr
}}
Here we have used that $V_{2}^{\vee }\cong V_{2}$, since it is a rank two
bundle of trivial determinant. Since $V_{4}^{\vee }$ has zero degree for any
polarization, all we must show is that the equivariant line subbundle 
${\cal O}_{\tilde{X}}(a_1\tau_1+a_2\tau_2+b\phi)$ has negative degree for the polarization we
are considering. If the compositions 
\eqn\hjrg{
{\cal O}_{\tilde{X}}(a_1\tau_1+a_2\tau_2+b\phi)\longrightarrow {\cal O}_{\tilde{X}}(\tau_1-\tau_2),
}
with each of the two direct factors on the right hand were both null, then
we would have a nonzero equivariant morphism 
\eqn\werew{
{\cal O}_{\tilde{X}}(a_1\tau_1+a_2\tau_2+b\phi)\longrightarrow \pi _{2}^{\ast }V_{2}(-\tau_1+\tau_2),
}
and these morphisms have been already analyzed in step one. Therefore, in
our situation we are necessarily in one of the following cases
\eqn\wedsdew{\matrix{
&& \cr
{\rm \, iii.1) \,} a_{1} &\leq & 1 {\rm \, and \, }a_{2}\leq -1{\rm \, and \, }b\leq
-{1\over 3}(a_{1}+a_{2}) \cr
{\rm \, iii.2) \,} a_{1} &\leq &-1 {\rm \, and \, }a_{2}\leq 0{\rm \, and \, }
b\leq 2-{1\over 3}(a_{1}+a_{2}) \cr
{\rm \, iii.3) \, } a_{1} &\leq & -1 {\rm \, and \, } a_{2}=1 {\rm \,  and \, } b\leq 
-{4\over 3}-{1\over 3}a_{1} \cr
{\rm \, iii.4) \, }a_{1} &\leq &-1{\rm \, and \, }a_{2}\leq 1{\rm \,  and \, }
b\leq -2-{1\over 3}(a_{1}+a_{2})\cr
}}

In case iii.1), the top instance ($a_{1}=1$ and $a_{2}=-1$ and $b=0$) 
would provide an essential
contradiction to what we want, if it ocurred, since no polarization giving 
${\cal O}_{\tilde{X}}(\tau_1-\tau_2)$ negative degree could give also negative degree to
the bundle ${\cal O}_{\tilde X}(-\tau_1+\tau_2)\oplus {\cal O}_{\tilde X}(-\tau_1+\tau_2)$ in the
presentation of $V_{4}$. Fortunately, this instance does not occur. Indeed,
in such a case the above morphism ${\cal O}_{\tilde{X}}(a_1\tau_1+a_2\tau_2+b\phi)
\to {\cal O}_{\tilde{X}}(\tau_1-\tau_2)$ would be isomorphic, thus spliting
the bottom sequence presenting $V_{3}$ in the diagram
\eqn\dedrfd{\matrix{
0 & \longrightarrow & {\cal O}_{\tilde X}(-\tau_1+\tau_2)\oplus {\cal O}_{\tilde X}(-\tau_1+\tau_2) & 
\longrightarrow & V_{4} & \longrightarrow & V_{2}(\tau_1-\tau_2) & \longrightarrow
& 0 \cr
&  & {\rm inclusion}\,\, \uparrow {\rm \, of \, one\, summand}&  & \uparrow &  & 
\uparrow {\rm \, id.} &  &  \cr 
0 & \longrightarrow & {\cal O}_{\tilde X}(-\tau_1+\tau_2) & \longrightarrow & V_{3} & 
\longrightarrow & V_{2}(\tau_1-\tau_2) & \longrightarrow & 0\cr
}}
in contradiction with the fact that the extension presenting $V_{4}$ has
been taken general, so with both of its components in the decomposition
\eqn\dferfde{\matrix{
&&{\rm Ext}^{1}(V_{2}(\tau_1-\tau_2),{\cal O}_{\tilde X}(-\tau_1+\tau_2)\oplus {\cal O}_{\tilde X}(-\tau_1+\tau_2)) \cr
&=&{\rm Ext}^{1}(V_{2}(\tau_1-\tau_2),{\cal O}_{\tilde X}(-\tau_1+\tau_2))\oplus {\rm Ext}^{1}(V_{2}(\tau_1-\tau_2), {\cal O}_{\tilde X}(-\tau_1+\tau_2)),\cr
}}
being nonzero. Therefore, the first case splits into three subcases:
\eqn\wedsdewe{\matrix{
&{\rm \, iii.1.a) \,}& a_{1} \leq  0 {\rm \, and \, }a_{2}\leq -1{\rm \, and \, }b\leq
-{1\over 3}(a_{1}+a_{2}) \cr
&{\rm \, iii.1.b) \,}& a_{1} \leq  1 {\rm \, and \, }a_{2}\leq -2{\rm \, and \, }b\leq
-{1\over 3}(a_{1}+a_{2}) \cr
&{\rm \, iii.1.c) \,}& a_{1} \leq  1 {\rm \, and \, }a_{2}\leq -1{\rm \, and \, }b\leq
-1-{1\over 3}(a_{1}+a_{2}) \cr
}}

Summing up, the vector bundle $V_4$ will then be stable if all the subsheaves that 
we have listed have negative degree. Recall that the degree
$d(x_{1},x_{2},y,a_{1},a_{2},b)$ is monotonous in $a_1$, $a_2$
and $b$, so in each case it is enough to check that it is negative
when these numbers take the maximum possible value.
Therefore, we get the following {\it sufficient} conditions for
a polarization to make $V_4$ stable:

\prop1
The vector bundle $V_{4}$ is equivariantly stable for any polarization 
${\cal O}_{\tilde{X}}(x_{1},x_{2},y)$ admitting equivariant structure 
(for instance, $x_{1},x_{2}$ multiple of 3) 
and making the number 
\eqn\dsoper{
d(x_{1},x_{2},y,a_{1},a_{2},b)
\,:=3(x_{1}+x_{2}+6y)(a_{1}x_{2}+a_{2}x_{1})+x_{1}x_{2}(3a_{1}+3a_{2}+18b),
}
negative for the following triples $(a_{1},a_{2},b)$ of integers 
\eqn\dsdere{\matrix{
&{\rm i.1)}& (-1,1,0)\cr
&{\rm i.2)}& (1,-2,7/3)        \cr
&{\rm i.3)}& (1,-1,-1)        \cr
&{\rm ii.1.b)}& (2,-3,0)        \cr
&{\rm ii.1.c)}& (2,-2,-1)        \cr
&{\rm iii.1.a)}& (0,-1,0)        \cr
&{\rm iii.2)}& (-1,0,5/3)        \cr
}}

\goodbreak\midinsert
\centerline{\epsfxsize 5truein\epsfbox{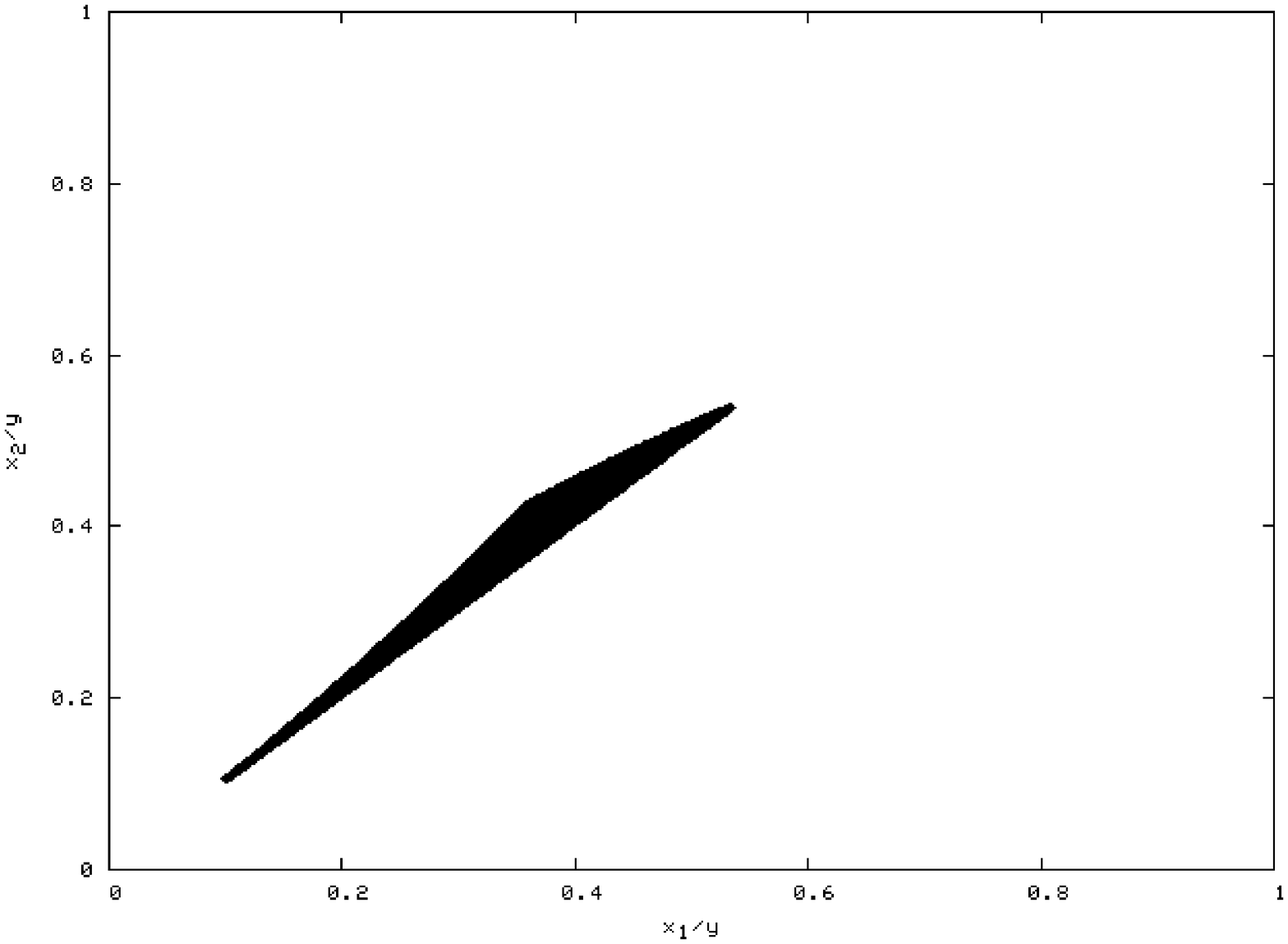}}
\leftskip 2pc \rightskip 2pc
\centerline{
\ninepoint\sl \baselineskip=8pt
{\bf Fig.1} Polarizations which make $V_4$ stable.}
\smallskip\endinsert

{\bf Remark} We have removed some cases which are redundant. For instance, case
i.4) corresponds to the point $(1,-1,-2)$, but this case 
is automatic once case i.3), 
corresponding to $(1,-1,-1)$, has been checked, since the degree function 
is monotonous in $a_1$, $a_2$ and $b$.

\bigskip
Using the proposition, it is easy to find examples of ample sheaves
which make $V_4$ stable. For instance, ${\cal O}_{\tilde X}(18\tau_1+21\tau_2+49\phi)$. In Figure 1 we have ploted the region of ample bundles which
satisfy the conditions of Proposition 1, and hence make stable the 
vector bundle $V_4$.

\bigskip
\noindent{\bf Acknowledgments}

\noindent{$ $}

We would like to thank E-D. Diaconescu, R. Donagi, A. Krause, G. Moore and G. Torroba
for useful discussions and correspondence, specially to 
M. E. Alonso for help with computer
algebra programs (Maple and Singular) and V. Braun for pointing out a computational mistake in a previous
version of this paper, fixing the mistake led us to a more explicit
form for Proposition 1.
S.L. thanks the 
UCM Department of Algebra for hospitality during his visit. 

\appendix{A}{ Action of the Mordell-Weil group on the homology }


The Mordell-Weil group $E(K)$, is defined adding sections fiberwise thanks to 
the group structure of an elliptic curve, once the zero section is fixed.  
More rigorously, we define $E(K)$ in terms of the short exact sequence 
\eqn\shortseq{
0\longrightarrow
T
\longrightarrow
H_{2}(B,\, \IZ)
\longrightarrow 
E(K)
\longrightarrow 0.
}
for certain subgroup $T$ in $H_{2}(B,\, \IZ)$.

For our elliptic surface, we know that the Mordell-Weil group
is isomorphic to $\IZ\oplus \IZ\oplus \IZ_{3}$ and is generated 
by the sections $\xi$, $\alpha_{B}\xi$ and $\eta$, thus we can express every section as
\eqn\decMW{
\boxplus x\xi\boxplus y\alpha_{B}\xi\boxplus z\eta\quad {\rm for}\,\, x,\, y\in \IZ\,\, 
{\rm and}\,\, z\in \IZ_{3}.
}
with $\boxplus x\xi$ (respectively $\boxplus y\alpha_{B}\xi$ and $\boxplus z\eta$) 
meaning 
$\boxplus x\xi=\underbrace{\xi\boxplus\xi\boxplus\dots \boxplus\xi}_{x}$.

Therefore, if $t_{a}:B\to B$ is the Mordell-Weil action of translating by the section $a$,
we have to determine the push forwards $(t_{\xi})_{\ast}$, $(t_{\alpha_{B}\xi})_{\ast}$ ,
$(t_{\eta})_{\ast}$ as maps $H_{2}(B)\to H_{2}(B)$, in order to express the homology class
of an arbitrary section as
\eqn\homide{
\big[\boxplus x\xi\boxplus y\alpha_{B}\xi\boxplus z\eta\big]=
(t_{\xi})_{\ast}^{x}\cdot (t_{\alpha_{B}\xi})_{\ast}^{y} \cdot 
(t_{\eta})_{\ast}^{z}\sigma
}
with $\sigma$ the zero section.

The push forwards $(t_{\xi})_{\ast}$, $(t_{\eta})_{\ast}$ and $(\alpha_{B})_{\ast}$ were already 
determined in \ellipticCY, using the quotient structure of the Mordell-Weil group on 
$H_{2}(B,\, \IZ)$ and computing intersection numbers with sections.
Here, we state their result, and derive $(t_{\alpha_{B}\xi})_{\ast}$ as
$(\alpha_{B})_{\ast}(t_{\xi})_{\ast}(\alpha_{B})_{\ast}^{-1}$, hence we have
\eqn\tabletwo{
(t_{\xi})_{\ast}
\cdot
\pmatrix{
    \sigma \cr F \cr
    \Theta_{1,1} \cr \Theta_{2,1} \cr \Theta_{3,1} \cr
    \Theta_{1,2} \cr \Theta_{2,2} \cr \Theta_{3,2} \cr
    \xi \cr \alpha_B \xi
    \cr 
}
=
\pmatrix{
0  & 0  & 0 & 0  & 0  & 0 & 0  & 0  &-1 & -1 \cr
0  & 1  & 0 & 0  & 1  & 0 & 1  & 0  & 0 & -1 \cr
0  & 0  & 1 & 0  & 0  & 0 & 0  & 0  & 0 & 0 \cr
0  & 0  & 0 & 0  & 0  & 0 & -1 & 0  & 1 & 0 \cr
0  & 0  & 0 & 0  & -1 & 0 & 0  & 1 & 0 & 1 \cr
0  & 0  & 0 & 0  & 0  & 1 & 0  & 0  & 0 & 0 \cr
0  & 0  & 0 & 1  & 0  & 0 & -1 & 0  & 0 & 0 \cr
0  & 0  & 0 & 0  & -1 & 0 & 0  & 0  & 1 & 1 \cr
 1 & 0  & 0 & 0  & 0  & 0 & 0  & 0  & 2 & 1 \cr
 0 & 0  & 0 & 0  & 0  & 0 & 0  & 0  & 0 & 1 \cr
}
\cdot
\pmatrix{
    \sigma \cr F \cr
    \Theta_{1,1} \cr \Theta_{2,1} \cr \Theta_{3,1} \cr
    \Theta_{1,2} \cr \Theta_{2,2} \cr \Theta_{3,2} \cr
    \xi \cr \alpha_B \xi
    \cr 
}}

\eqn\tablethr{
(t_{\alpha_{B}\xi})_{\ast}
\cdot
\pmatrix{
    \sigma \cr F \cr
    \Theta_{1,1} \cr \Theta_{2,1} \cr \Theta_{3,1} \cr
    \Theta_{1,2} \cr \Theta_{2,2} \cr \Theta_{3,2} \cr
    \xi \cr \alpha_B \xi
    \cr 
}
=
\pmatrix{
0 & 0 & 0 & 0& 0 & 0 & 0 & 0 &-1 & -1\cr
0 & 1 & 1 & 0& 0 & 0 & 0 & 1 &-1 & 0\cr
0 & 0 &-1 & 0& 0 & 1 & 0 & 0 & 0 & 0\cr
0 & 0 & 0 & 1& 0 & 0 & 0 & 0 & 0 & 0\cr
0 & 0 & 0 & 0& 0 & 0 & 0 &-1 & 1 & 1\cr
0 & 0 &-1 & 0& 0 & 0 & 0 & 0 & 0 & 1\cr
0 & 0 & 0 & 0& 0 & 0 & 1 & 0 & 0 & 0\cr
0 & 0 & 0 & 0& 1 & 0 & 0 &-1 & 1 & 0\cr
0 & 0 & 0 & 0& 0 & 0 & 0 & 0 & 1 & 0\cr
1 & 0 & 0 & 0& 0 & 0 & 0 & 0 & 1 & 2\cr
}
\cdot
\pmatrix{
    \sigma \cr F \cr
    \Theta_{1,1} \cr \Theta_{2,1} \cr \Theta_{3,1} \cr
    \Theta_{1,2} \cr \Theta_{2,2} \cr \Theta_{3,2} \cr
    \xi \cr \alpha_B \xi
    \cr 
}}

\eqn\tablefou{
(t_{\eta})_{\ast}
\cdot
\pmatrix{
    \sigma \cr F \cr
    \Theta_{1,1} \cr \Theta_{2,1} \cr \Theta_{3,1} \cr
    \Theta_{1,2} \cr \Theta_{2,2} \cr \Theta_{3,2} \cr
    \xi \cr \alpha_B \xi
    \cr 
}
=
\pmatrix{
1    & 0 & 0     & 0     & 0    & 0     & 0     & 0     & 0   & 0   \cr
1    & 1 & 0     & 0     & 0    & 1     & 1     & 1     & 0   & 0   \cr
-2/3 & 0 & 0     & 0     & 0    & -1    & 0     & 0     &-2/3 & 1/3 \cr
-2/3 & 0 & 0     & 0     & 0    & 0     & -1    & 0     & 1/3 & -2/3\cr
-2/3 & 0 & 0     & 0     & 0    & 0     & 0     & -1    & 1/3 & 1/3 \cr
-1/3 & 0 &  1    & 0     & 0    & -1    & 0     & 0     &-1/3 & 2/3 \cr
-1/3 & 0 & 0     &  1    & 0    & 0     & -1    & 0     &-1/3 & -1/3\cr
-1/3 & 0 & 0     & 0     & 1    & 0     & 0     & -1    & 2/3 & -1/3\cr
0    & 0 & 0     & 0     & 0    & 0     & 0     & 0     & 1   & 0   \cr
0    & 0 & 0     & 0     & 0    & 0     & 0     & 0     & 0   & 1   \cr
}
\cdot
\pmatrix{
    \sigma \cr F \cr
    \Theta_{1,1} \cr \Theta_{2,1} \cr \Theta_{3,1} \cr
    \Theta_{1,2} \cr \Theta_{2,2} \cr \Theta_{3,2} \cr
    \xi \cr \alpha_B \xi
    \cr 
}}
Another way of looking at these three matrices is as 
generators of the representation of the Mordell-Weil group
in ${\rm End}\,\big( H_{2}(B,\,\IZ)\big)$. The commutation
relations $[(t_{\xi})_{\ast},\,(t_{\alpha_{B}\xi})_{\ast}]=0$,
$[(t_{\xi})_{\ast},\,(t_{\eta})_{\ast}]=0$,
$[(t_{\eta})_{\ast},\,(t_{\alpha_{B}\xi})_{\ast}]=0$ are 
obeyed and the torsion generator 
$(t_{\eta})_{\ast}$, verifies $(t_{\eta})_{\ast}^{3}={\bf 1}$
as expected.

\noindent{\rm Thus}, expanding the equation 
\eqn\homide{
\big[\boxplus x\xi\boxplus y\alpha_{B}\xi\boxplus z\eta\big]=
(t_{\xi})_{\ast}^{x}\cdot (t_{\alpha_{B}\xi})_{\ast}^{y} \cdot 
(t_{\eta})_{\ast}^{z}\sigma
}
for the homology classes of the sections,
gives us the following list\foot{It can be proven to hold by using induction.}:

\noindent{\rm If}
\eqn\firsection{\pmatrix{x\cr y\cr z\cr} \equiv 
\pmatrix{0\cr 0\cr 0\cr}
({\rm mod}\, 3)\quad
{\rm or} \quad
\pmatrix{2\cr 1\cr 1\cr}
({\rm mod}\, 3)\quad
{\rm or} \quad
\pmatrix{1\cr 2\cr 2\cr}
({\rm mod}\, 3)
}
then

\noindent{
$\big[\boxplus x\xi\boxplus y\alpha_{B}\xi\boxplus z\eta\big]=
(1-x-y)\sigma + 
(1/3x^2 + 1/3y^2 - 1/3xy -x -y)F+
1/3y\Theta_{1,1}+
2/3x\Theta_{2,1}+
(1/3x+2/3y)\Theta_{3,1}+
2/3y\Theta_{1,2}+
1/3x\Theta_{2,2}+
(2/3x+1/3y)\Theta_{3,2}+
x\xi +
y\alpha_{B}\xi.
$}

\noindent{\rm If}
\eqn\firsection{\pmatrix{x\cr y\cr z\cr} \equiv 
\pmatrix{1\cr 2\cr 0\cr}
({\rm mod}\, 3)\quad
{\rm or} \quad
\pmatrix{0\cr 0\cr 1\cr}
({\rm mod}\, 3)\quad
{\rm or} \quad
\pmatrix{2\cr 1\cr 2\cr}
({\rm mod}\, 3)
}
then

\noindent{
$\big[\boxplus x\xi\boxplus y\alpha_{B}\xi\boxplus z\eta\big]=
(1-x-y)\sigma + 
(1/3x^2 + 1/3y^2 - 1/3xy -x -y+1)F+
(1/3y-2/3)\Theta_{1,1}+
(2/3x-2/3)\Theta_{2,1}+
(1/3x+2/3y-2/3)\Theta_{3,1}+
(2/3y-1/3)\Theta_{1,2}+
(1/3x-1/3)\Theta_{2,2}+
(2/3x+1/3y-1/3)\Theta_{3,2}+
x\xi +
y\alpha_{B}\xi.
$}

\noindent{\rm If}
\eqn\firsection{\pmatrix{x\cr y\cr z\cr} \equiv 
\pmatrix{2\cr 1\cr 0\cr}
({\rm mod}\, 3)\quad
{\rm or} \quad
\pmatrix{1\cr 2\cr 1\cr}
({\rm mod}\, 3)\quad
{\rm or} \quad
\pmatrix{0\cr 0\cr 2\cr}
({\rm mod}\, 3)
}
then

\noindent{
$\big[\boxplus x\xi\boxplus y\alpha_{B}\xi\boxplus z\eta\big]=
(1-x-y)\sigma + 
(1/3x^2 + 1/3y^2 - 1/3xy -x -y+1)F+
(1/3y-1/3)\Theta_{1,1}+
(2/3x-1/3)\Theta_{2,1}+
(1/3x+2/3y-1/3)\Theta_{3,1}+
(2/3y-2/3)\Theta_{1,2}+
(1/3x-2/3)\Theta_{2,2}+
(2/3x+1/3y-2/3)\Theta_{3,2}+
x\xi +
y\alpha_{B}\xi.
$}

\noindent{\rm If}
\eqn\firsection{\pmatrix{x\cr y\cr z\cr} \equiv 
\pmatrix{0\cr 1\cr 0\cr}
({\rm mod}\, 3)\quad
{\rm or} \quad
\pmatrix{2\cr 2\cr 1\cr}
({\rm mod}\, 3)\quad
{\rm or} \quad
\pmatrix{1\cr 0\cr 2\cr}
({\rm mod}\, 3)
}
then

\noindent{
$\big[\boxplus x\xi\boxplus y\alpha_{B}\xi\boxplus z\eta\big]=
(1-x-y)\sigma + 
(1/3x^2 + 1/3y^2 - 1/3xy -x -y+2/3)F+
(1/3y-1/3)\Theta_{1,1}+
2/3x\Theta_{2,1}+
(1/3x+2/3y-2/3)\Theta_{3,1}+
(2/3y-2/3)\Theta_{1,2}+
1/3x\Theta_{2,2}+
(2/3x+1/3y-1/3)\Theta_{3,2}+
x\xi +
y\alpha_{B}\xi.
$}

\noindent{\rm If}
\eqn\firsection{\pmatrix{x\cr y\cr z\cr} \equiv 
\pmatrix{1\cr 0\cr 0\cr}
({\rm mod}\, 3)\quad
{\rm or} \quad
\pmatrix{0\cr 1\cr 1\cr}
({\rm mod}\, 3)\quad
{\rm or} \quad
\pmatrix{2\cr 2\cr 2\cr}
({\rm mod}\, 3)
}
then

\noindent{
$\big[\boxplus x\xi\boxplus y\alpha_{B}\xi\boxplus z\eta\big]=
(1-x-y)\sigma + 
(1/3x^2 + 1/3y^2 - 1/3xy -x -y+2/3)F+
1/3y\Theta_{1,1}+
(2/3x-2/3)\Theta_{2,1}+
(1/3x+2/3y-1/3)\Theta_{3,1}+
2/3y\Theta_{1,2}+
(1/3x-1/3)\Theta_{2,2}+
(2/3x+1/3y-2/3)\Theta_{3,2}+
x\xi +
y\alpha_{B}\xi.
$}

\noindent{\rm If}
\eqn\firsection{\pmatrix{x\cr y\cr z\cr} \equiv 
\pmatrix{2\cr 2\cr 0\cr}
({\rm mod}\, 3)\quad
{\rm or} \quad
\pmatrix{1\cr 0\cr 1\cr}
({\rm mod}\, 3)\quad
{\rm or} \quad
\pmatrix{0\cr 1\cr 2\cr}
({\rm mod}\, 3)
}
then

\noindent{
$\big[\boxplus x\xi\boxplus y\alpha_{B}\xi\boxplus z\eta\big]=
(1-x-y)\sigma + 
(1/3x^2 + 1/3y^2 - 1/3xy -x -y+2/3)F+
(1/3y-2/3)\Theta_{1,1}+
(2/3x-1/3)\Theta_{2,1}+
(1/3x+2/3y)\Theta_{3,1}+
(2/3y-1/3)\Theta_{1,2}+
(1/3x-2/3)\Theta_{2,2}+
(2/3x+1/3y)\Theta_{3,2}+
x\xi +
y\alpha_{B}\xi.
$}

\noindent{\rm If}
\eqn\firsection{\pmatrix{x\cr y\cr z\cr} \equiv 
\pmatrix{0\cr 2\cr 0\cr}
({\rm mod}\, 3)\quad
{\rm or} \quad
\pmatrix{2\cr 0\cr 1\cr}
({\rm mod}\, 3)\quad
{\rm or} \quad
\pmatrix{1\cr 1\cr 2\cr}
({\rm mod}\, 3)
}
then

\noindent{
$\big[\boxplus x\xi\boxplus y\alpha_{B}\xi\boxplus z\eta\big]=
(1-x-y)\sigma + 
(1/3x^2 + 1/3y^2 - 1/3xy -x -y+2/3)F+
(1/3y-2/3)\Theta_{1,1}+
2/3x\Theta_{2,1}+
(1/3x+2/3y-1/3)\Theta_{3,1}+
(2/3y-1/3)\Theta_{1,2}+
1/3x\Theta_{2,2}+
(2/3x+1/3y-2/3)\Theta_{3,2}+
x\xi +
y\alpha_{B}\xi.
$}

\noindent{\rm If} 
\eqn\firsection{\pmatrix{x\cr y\cr z\cr} \equiv 
\pmatrix{1\cr 1\cr 0\cr}
({\rm mod}\, 3)\quad
{\rm or} \quad
\pmatrix{0\cr 2\cr 1\cr}
({\rm mod}\, 3)\quad
{\rm or} \quad
\pmatrix{2\cr 0\cr 2\cr}
({\rm mod}\, 3)
}
then

\noindent{
$\big[\boxplus x\xi\boxplus y\alpha_{B}\xi\boxplus z\eta\big]=
(1-x-y)\sigma + 
(1/3x^2 + 1/3y^2 - 1/3xy -x -y+2/3)F+
(1/3y-1/3)\Theta_{1,1}+
(2/3x-2/3)\Theta_{2,1}+
(1/3x+2/3y)\Theta_{3,1}+
(2/3y-2/3)\Theta_{1,2}+
(1/3x-1/3)\Theta_{2,2}+
(2/3x+1/3y)\Theta_{3,2}+
x\xi +
y\alpha_{B}\xi.
$}

\noindent{\rm If}
\eqn\firsection{\pmatrix{x\cr y\cr z\cr} \equiv 
\pmatrix{2\cr 0\cr 0\cr}
({\rm mod}\, 3)\quad
{\rm or} \quad
\pmatrix{1\cr 1\cr 1\cr}
({\rm mod}\, 3)\quad
{\rm or} \quad
\pmatrix{0\cr 2\cr 2\cr}
({\rm mod}\, 3)
}
then

\noindent{
$\big[\boxplus x\xi\boxplus y\alpha_{B}\xi\boxplus z\eta\big]=
(1-x-y)\sigma + 
(1/3x^2 + 1/3y^2 - 1/3xy -x -y+2/3)F+
1/3y\Theta_{1,1}+
(2/3x-1/3)\Theta_{2,1}+
(1/3x+2/3y-2/3)\Theta_{3,1}+
2/3y\Theta_{1,2}+
(1/3x-2/3)\Theta_{2,2}+
(2/3x+1/3y-1/3)\Theta_{3,2}+
x\xi +
y\alpha_{B}\xi.
$}

\listrefs
\end